\documentclass[clusterscriptaddress,showpacs,aps,pre,twocolumn,nofootinbib,10]{revtex4-1}
\usepackage{epsfig}
\usepackage{bm}
\usepackage{amsmath,slashed}
\usepackage{subfigure}
\usepackage{graphicx}
\usepackage{epstopdf}
\usepackage[pdftex]{color}
\usepackage{hyperref}
\hypersetup{
    colorlinks=true,       
    linkcolor=cyan,          
    citecolor=magenta,        
    filecolor=magenta,      
    urlcolor=cyan,           
    runcolor=cyan
}

\newcommand{\beq}{\begin{equation}}
\newcommand{\eeq}{\end{equation}}
\newcommand{\bea}{\begin{eqnarray}}
\newcommand{\eea}{\end{eqnarray}}
\newcommand{\bes} {\begin{subequations}}
\newcommand{\ees} {\end{subequations}}

\newcommand{\e}{{\text e}}

\newcommand{\HI}{H}

\newcommand{\ignore}[1]{}

\usepackage{algorithm}
\usepackage{algpseudocode}
\begin{document}

\title{Solving Spin Glasses with Optimized Trees of Clustered Spins}

\author{Itay Hen}
\affiliation{Information Sciences Institute, University of Southern California, Marina del Rey, CA 90292, USA}
\affiliation{Department of Physics and Astronomy and Center for Quantum Information Science \& Technology, University of Southern California, Los Angeles, California 90089, USA}
\email{itayhen@isi.edu}


\begin{abstract}
We present an algorithm for the optimization and thermal equilibration of spin glasses---or more generally, cost functions of the Ising form $H=\sum_{\langle i j\rangle} J_{ij} s_i s_j + \sum_i h_i s_i$, defined on graphs with arbitrary connectivity. The algorithm consists of two repeated steps: i) the optimized construction of a random tree of spin clusters on the input problem graph, and ii) the thermal sampling of the generated tree. The randomly generated trees are constructed so as to optimize a balance between the size of the tree and the complexity required to draw Boltzmann samples from it.   
We benchmark the algorithm on several classes of problems and demonstrate its advantages over existing approaches.  
\end{abstract}

\maketitle

\section{Introduction}
Many problems of theoretical and practical relevance consist of searching for the global minimum of intricate cost functions defined over discrete configuration spaces. These so-called combinatorial optimization problems are not only notoriously hard to solve, but are also ubiquitous, appearing in a wide
range of diverse fields such as machine learning, materials design, software verification, flight-traffic control, constraint satisfaction problems and logistics, to name a few diverse examples~\cite{wolsey,papadimitriou2013combinatorial}. It is no surprise then that the design of fast and practical algorithms to solve these problems is an important challenge in many areas of science and technology.

A class of problems that is known to be particularly challenging to solve is that of 
\emph{spin-glasses\/}~\cite{young:98}---disordered, frustrated spin systems, so intricate
that specialized hardware has been built to simulate
them~\cite{janus:08,janus:09,janus:14}. The spin glass, or Ising model,
cost function is usually written as
\beq\label{eq:Hgeneral} 
\HI=\sum_{\langle ij\rangle} J_{ij} s_i
s_j + \sum_i h_i s_i
\eeq
defined over a set of binary variables, the Ising spins $\{s_i=\pm 1\}$. The spins sit at the vertices of an interaction
graph defined by the sets of given parameters $\{J_{ij}\}$ and
$\{h_i\}$.  Here, $\sum_{\langle ij\rangle}$ denotes a sum over all the edges
of the connectivity graph of the problem. Finding the minimal cost and spin
assignments that produce it, or using physics terminology, the ground state energy and 
corresponding spin configurations for problem instances belonging to the above model, is generally known to be a
combinatorially intractable (or, NP-hard) optimization
problem~\cite{barahona:82}. Furthermore, a significant portion of real-life hard discrete optimization problems can be cast as spin glasses~\cite{lucasIsing}, 
making the development of novel practical algorithms with significantly shorter runtimes and/or  superior 
scaling with problem size  the holy grail of optimization.  

The problem of finding the lowest costs of spin-glass Hamiltonians, Eq.~(\ref{eq:Hgeneral}),
can be and is often generalized to an analogous problem in statistical mechanics, where one is asked to 
draw spin configurations from the Boltzmann distribution $\mathcal{B}(T)$ of that cost function for a given temperature $T$---or expressed differently, to sample the configuration space of the problem such that the probability of drawing any given spin configuration ${\mathbf s}=(s_1,\ldots, s_N)$ is proportional to its Boltzmann weight 
$W_{\mathbf s} = \e^{-\beta E_{\mathbf s}}$ where $\beta=1/T$ and $E_{\mathbf s}$ is the cost associated with the configuration. In the zero temperature limit $T \to 0$, Boltzmann sampling reduces to the standard optimization variant of the problem wherein only ground state assignments have nonzero weights. 

The Boltzmann sampling of spin glasses at nonzero temperatures has numerous immediate practical applications in of itself in various areas (e.g., in machine learning~\cite{rbm}). Furthermore, and perhaps more importantly, Boltzmann sampling is known to serve as a useful subroutine in optimization algorithms where only the absolute lowest cost is sought.
The observation that the thermalization, or equilibration, of intricate cost functions at sequentially decreasing temperatures is beneficial for finding optimal assignments  
was first made by Kirkpatrick \emph{et al.}~\cite{kirkpatrick:83} who, based on this observation, pioneered the now-celebrated approach commonly referred to as `simulated annealing' (SA).
Today, the methods of choice for studying general
spin-glass problems are variations of SA.  One scheme is  `population annealing'~\cite{populationAnnealing} in which simulated annealing is combined with Boltzmann weighted differential reproduction within a population of replicas to sample equilibrium states. Another technique, that we describe in more detail in Sec.~\ref{sec:PT}, is 
`parallel tempering' (PT)~\cite{hukushima:96,marinari:98b} in which multiple copies of the problem are equilibrated in parallel at different temperatures and spin configurations at adjacent temperatures are regularly swapped. 

At the most basic level, the equilibration of spin glasses at any given temperature $T$ consists of the repeated updating of a stored spin configuration ${\mathbf s}$ where usually one calculates the energy change $\Delta E=E_{{\mathbf s}_{\text{new}}}-E_{\mathbf s}$ associated with the flipping of a single spin while keeping the values of all other spins fixed and accepting the move with a probability that satisfies a detailed balance, namely, 
\hbox{$P_{{\mathbf s} \to {\mathbf s}_{\text{new}}}/P_{{\mathbf s}_{\text{new}}\to {\mathbf s}} =W_{{\mathbf s}_{\text{new}}}/W_{\mathbf s}$}. 
Here, ${\mathbf s}$ denotes the current spin configuration and ${\mathbf s}_{\text{new}}$ is the proposed new configuration.
The above condition is sufficient to ensure the eventual proper thermalization of the system. 
For spin glasses, updates consisting of single spin-flips (SSFs) are often found to be grossly inefficient. This is mainly because the system often finds itself trapped at local minima (or metastable states) of the cost function, usually surrounded by high energy barriers. The probability for a spin configuration to `jump over' such barriers is exponentially suppressed in the barrier height $\Delta E$. 
On the other hand, moves consisting of the thermal sampling of larger subsystems, i.e., multiple spin-flips, which correspond to potentially higher jumps, are generally very costly to implement, as the complexity of calculating the energy changes and Boltzmann weights associated with multiple spin-flips grows exponentially with the size of the spin cluster that is to be flipped.  

Certain subgraph structures, however, can nonetheless be sampled, or optimized, more efficiently. The best example is \emph{trees}---subgraphs that do not contain any cycles. Trees are known to give rise to equilibration techniques, such as belief propagation~\cite{beliefBook}, which allow for a more efficient Boltzmann sampling, or ground state optimization, than that of single spin updates~\cite{selby,hamze:04, PhysRevB.89.214421}. The computational complexity associated with the Boltzmann sampling of trees of spins scales only linearly 
with their size.  However, optimization of spin glasses via trees of single spins (TSSs) induced on input problem connectivity graphs is very often found to be inefficient compared to single spin-flip techniques. In part, this is because arbitrarily connected spin glass graphs may contain many short cycles, and thus severely restrict the size of the induced trees. Moreover, the computational overhead associated with the generation of random trees on the instance connectivity graph plays a role in diminishing the scaling advantage of using them~\cite{PhysRevB.89.214421}.

In this work we propose a novel algorithm for the thermalization and optimization of spin glasses. The algorithm aims to adequately address the pitfalls of existing approaches discussed above---namely, the inefficiency of single spin-flips and the impracticality that is often associated with the sampling of trees of single spins.  The algorithm we propose here is based on heat-bath updates performed on \emph{optimized trees of spin clusters} induced on the input problem graph. The algorithm consists of two basic repeated steps. In the first step, a tree of clusters of spins is induced on the spin glass graph in a manner which optimizes a certain balance between the size of the subgraph that is to be thermalized and the complexity of doing so. As a next step, 
a configuration from the Boltzmann distribution $\mathcal{B}(T)$ of the tree is drawn (while keeping the spins not belonging to the tree fixed). As we demonstrate, a methodical grouping together of spins into clusters generally allows for the construction of much larger trees on the input graph---a process which in turn allows for the efficient thermalization of large subgraphs, thereby generally speeding up the entire equilibration process. 

This paper is organized as follows. In Sec.~\ref{sec:algorithm} we discuss the two steps of the algorithm, namely, the processes for generation and subsequent sampling of trees of spin clusters. We also illustrate how one can use the complexity of sampling a tree of clustered spins to construct a figure of merit with which optimal subgraph structures can be generated on the input problem graph and then thermalized. In Sec.~\ref{sec:results} we present the results of several benchmarking tests comparing the performance of the proposed algorithm, as an optimizer as well as for thermal equilibration, to existing techniques---specifically, single spin-flip and random single spin tree methods. Section~\ref{sec:discussion} is devoted to a discussion of the results and the applicability of the method to optimization problems of practical relevance. 

\section{Thermal annealing with optimized trees of spin clusters~\label{sec:algorithm}}

For reasons that will become clear shortly, we begin our description of the algorithm with the second of its two steps, namely, the subroutine for the Boltzmann sampling of a tree of clustered spins. The complexity of this subroutine, which we calculate here, will play a part in the design of the first step of the algorithm, within which the tree structures that are to be thermalized in the second step are constructed. 

\subsection{Boltzmann sampling of a tree of clustered spins\label{sec:clusterIsing}}

\begin{figure}[ht] 
\subfigure[]{\includegraphics[angle=270,trim={1cm 1cm 16cm 1cm},clip,width=0.65\columnwidth]{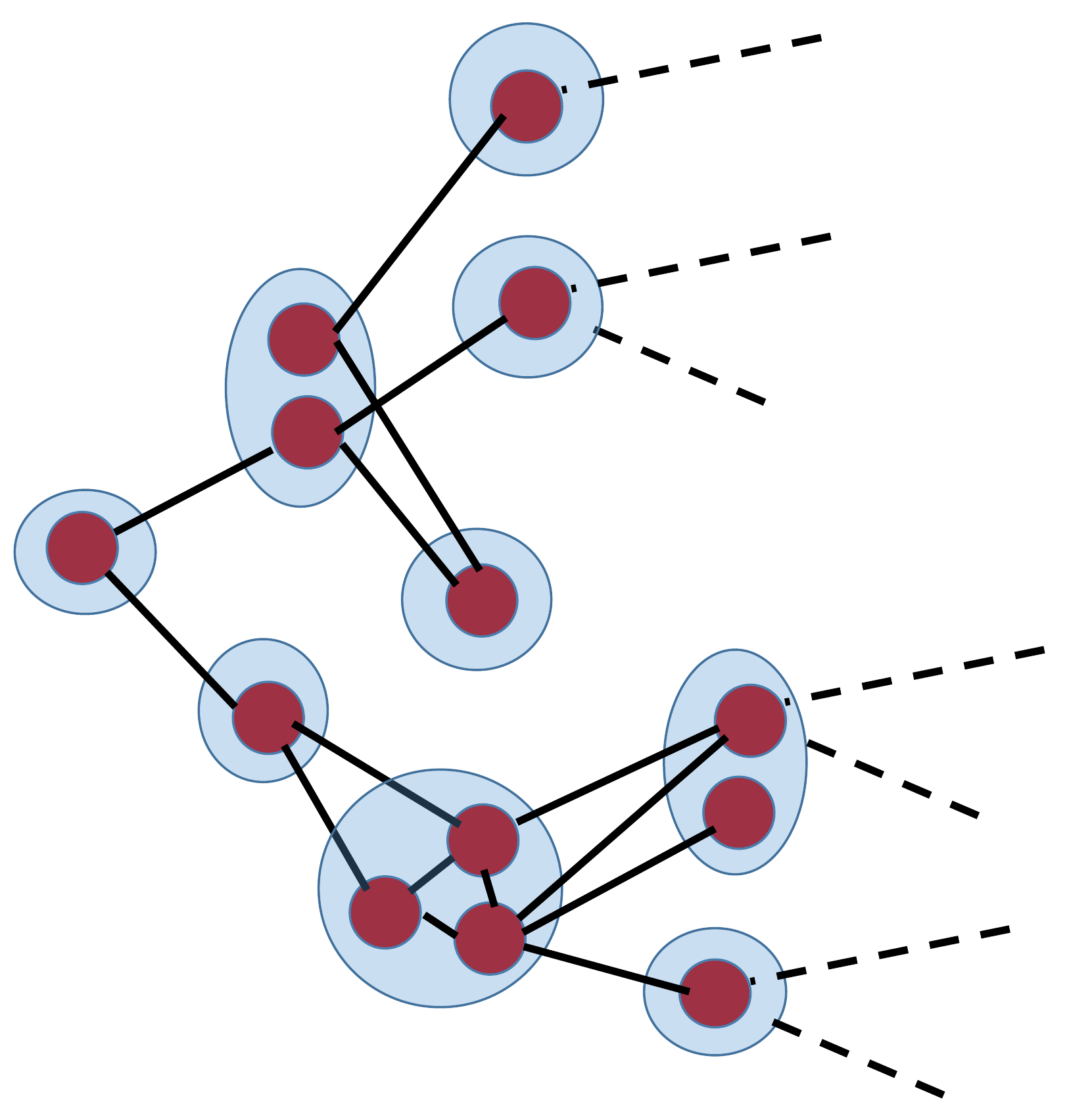}}
\subfigure[]{\includegraphics[width=0.99\columnwidth]{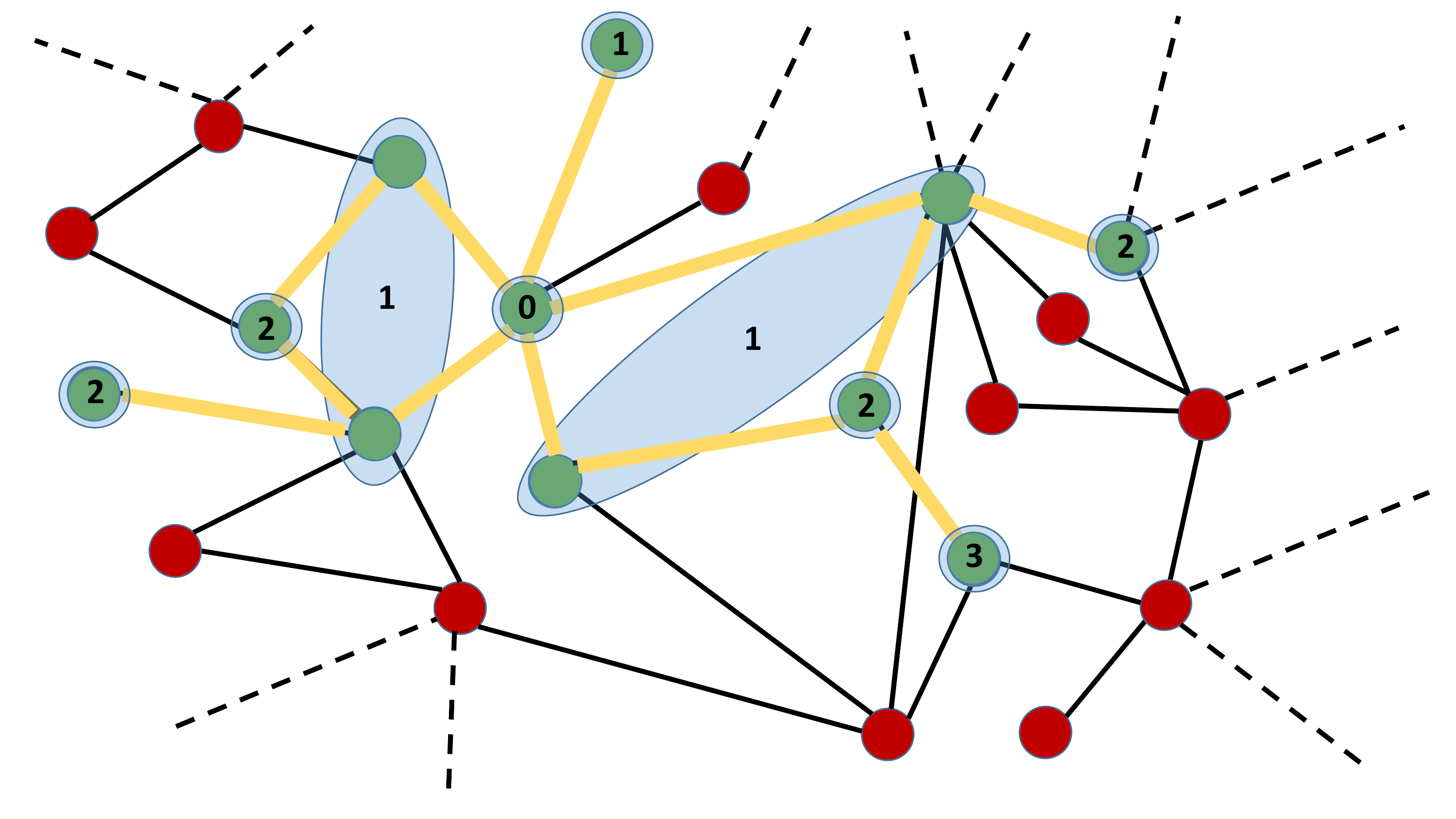}}
   \caption{{\bf (a) A tree of clustered spins.} Each tree node is a group, or a cluster, of spins that may or may not have edges between them. Edges between two tree nodes appear if and only if there are Ising interactions between their spins. {\bf (b) A tree of clustered spins embedded in a larger input problem graph.} The tree nodes are marked with light blue ellipses encircling individual spins in the tree (green circles). Red circles are spins not belonging to the tree. The numerical labels denote the distance of a node to the tree root (labeled by $0$). }
   \label{fig:trees}
\end{figure}

Given an input problem of the Ising type, Eq.~(\ref{eq:Hgeneral}), we define a \emph{cluster} as a set of spins of the input problem that may or may not be coupled among themselves. A \emph{tree of spin clusters} (or a TOSC) is a subgraph of the input problem graph that has the structure of a tree---that is, each node on the tree is a cluster of spins, and edges between neighboring nodes exist if and only if Ising interactions between their respective spins exist. Figure~\ref{fig:trees}(a) illustrates a tree of clustered spins. 

We now discuss an algorithm for drawing a spin configuration on a TOSC with Boltzmann probabilities of a given inverse temperature $\beta$. Fixing the values of all the input-problem spins not belonging to the tree, interactions with outside spins will henceforth be treated as fixed external fields.
Figure~\ref{fig:trees}(b) illustrates a tree of clustered spins embedded in a larger input problem graph.   
For what follows, we shall denote tree nodes (i.e., spin clusters) or configurations thereof by \hbox{$\mathbf{s}_i=\{s_{i(1)},s_{i(2)},\ldots\}$} where $i$ labels the node, and $s_{i(k)}$ denotes the individual spins comprising the node.
 
For brevity, we shall denote the combined interaction energy between spins in neighboring nodes $i$ and $j$ as
\beq
E_{\text{inter}}(\mathbf{s}_i,\mathbf{s}_j)=\sum_{k,k'} J_{i(k),j(k')} s_{i(k)} s_{j(k')}\,,
\eeq
and the energy associated with the interaction of spins within a node, combined with the energy associated with external fields (including the interaction energy with fixed spins not in the tree), as
\beq
E_{\text{self}}(\mathbf{s}_i)=\sum_k h_{i(k)} s_{i(k)} +\sum_{k<k'} J_{i(k),i(k')} s_{i(k)} s_{i(k')}\,.
\eeq
We consider now a tree of clustered spins with a root node $\mathbf{s}_i$. Let us denote the set of spins comprising the tree by $\mathbf{t}_{\mathbf{s}_i}$ and configurations of spins on the tree by $t_{\mathbf{s}_i}$.   
Denoting the set of children nodes of $\mathbf{s}_i$ by $c(\mathbf{s}_i)$, and $\mathbf{s}_j \in c(\mathbf{s}_i)$ as the children nodes, we have $\mathbf{t}_{\mathbf{s}_i} = \{\mathbf{s}_i\} \cup_j \mathbf{t}_{\mathbf{s}_j}$;
i.e., the set of clustered spins forming a tree is the union of the root node with all the \hbox{(sub-)tree} sets whose root nodes are $\mathbf{s}_j$. 
The energy $E(t_{\mathbf{s}_i})$ of any given tree configuration $t_{\mathbf{s}_i}$ can thus be written as
\bea
E(t_{\mathbf{s}_i}) &=& E_{\text{self}}(\mathbf{s}_i)+ \sum_{\mathbf{s}_j\in c(\mathbf{s}_i)} E_{\text{inter}}(\mathbf{s}_i,\mathbf{s}_j)  \\\nonumber
&+& \sum_{\mathbf{s}_j\in c(\mathbf{s}_i)} E(t_{\mathbf{s}_j}) \,.
\eea
From the above definition, it follows that the weight associated with a configuration $t_{\mathbf{s}_i}$ can be written as 
\beq\label{eq:Wt}
W(t_{\mathbf{s}_i}) =\e^{-\beta E_{\text{self}}(\mathbf{s}_i)} \prod_j \left[ \e^{-\beta E_{\text{inter}}(\mathbf{s}_i,\mathbf{s}_j)} W(t_{\mathbf{s}_j}) \right]\,,
\eeq
where the goal of the algorithm is to generate a spin configuration $t_{\mathbf{s}_i}$ with a probability 
that is proportional to 
$W(t_{\mathbf{s}_i})$; that is,
\beq
P(t_{\mathbf{s}_i}) = \frac{W(t_{\mathbf{s}_i})}{\sum_{t_{\mathbf{s}_i}} W(t_{\mathbf{s}_i})} \,.
\eeq
To generate a thermal configuration over the tree $\mathbf{t}_{\mathbf{s}_i}$ with a Boltzmann probability, we will first calculate the probability $P(\mathbf{s}_i)$ of assigning a specific value to $\mathbf{s}_i$ (equivalently, specific values to the spins of  $\mathbf{s}_i$).
After $\mathbf{s}_i$ is fixed, the problem immediately decouples to the thermalization of a set of disconnected (sub-)trees $\mathbf{t}_{\mathbf{s}_j}$ for which the process will be repeated. 

To calculate the probability $P(\mathbf{s}_i)$  of assigning a specific value $\mathbf{s}_i$ to the root node, we shall use the relation 
\hbox{$P(A \cap B) = P(A \vert B) P(B)$}, from which it follows that the probability for drawing a configuration of spins on the tree can be written as the product 
\beq
P(t_{{\mathbf{s}}_{i}})=P(\mathbf{s}_i,t_{\slashed{\mathbf{s}}_{i}})=P(t_{\slashed{\mathbf{s}}_{i}} \vert \mathbf{s}_i) P(\mathbf{s}_i) \,.
\eeq
where $t_{\slashed{\mathbf{s}}_i}$ is any configuration of spins over the set $\mathbf{t}_{\mathbf{s}_i}$ with the value of $\mathbf{s}_i$ fixed. 
Rearranging, we find that:
\beq
P(\mathbf{s}_i)=\frac{P(\mathbf{s}_i,t_{\slashed{\mathbf{s}}_{i}})} {P(t_{\slashed{\mathbf{s}}_{i}}| \mathbf{s}_i)}=\frac{\sum_{t_{\slashed{\mathbf{s}}_i}} W(t_{\mathbf{s}_i})}{\sum_{t_{\mathbf{s}_i}} W(t_{\mathbf{s}_i})}
 \,,
\eeq
where $\sum_{t_{\mathbf{s}_i}}$ denotes a sum over all spin configurations on the set $\mathbf{t}_{\mathbf{s}_i}$, and 
$\sum_{t_{\slashed{\mathbf{s}}_i}}$
denotes a sum over all configurations over the set $t_{\mathbf{s}_i}$ with the value of the root node set to $\mathbf{s}_i$. Rewriting $\sum_{t_{\mathbf{s}_i}}=\sum_{{\mathbf{s}_i}}\sum_{t_{\slashed{\mathbf{s}}_i}}$, the above becomes:
\beq
P(\mathbf{s}_i)=\frac{W(\mathbf{s}_i)}{\sum_{\mathbf{s}'_i} W(\mathbf{s}'_i)}\,,
\eeq
where we have defined the weight of a node $\mathbf{s}_i$ as: 
\bea
W(\mathbf{s}_i)&=&\sum_{t_{\slashed{\mathbf{s}}_i}} W(t_{\mathbf{s}_i})
 \\\nonumber
&=& \e^{-\beta E_{\text{self}}(\mathbf{s}_i)} \prod_j \left[ \sum_{t_{\mathbf{s}_j}} 
\e^{-\beta E_{\text{inter}}(\mathbf{s}_i,\mathbf{s}_j)}
W(t_{\mathbf{s}_j}) \right]\,.
 \eea
We thus find that the probability of assigning a specific value to any clustered spin $\mathbf{s}_i$ can be calculated from its weight, which is obtained from the weight of the tree of which $\mathbf{s}_i$ is the root. 
After fixing the root node $\mathbf{s}_i$, the probabilities of its children $\mathbf{s}_j$ can similarly be determined and their values subsequently fixed.

To assign a configuration to all nodes in a tree, we follow these two steps: i) Advancing from the tree leaves (that is, from nodes with no children) to root, we calculate the weights of all the possible node configurations. The calculation of the weight of a node, Eq.~(\ref{eq:Wt}), is determined by its self-energy, the interaction energy with its children and their weights.
ii) Having obtained the weights of the various nodes, we begin assigning configurations to the tree nodes, starting from the root node advancing toward the tree leaves. Assigning a configuration to a node becomes possible once the configuration of its parent node has been fixed, at which point the node becomes a root node for a subtree, and the node's self energy is amended to include its interaction with its (now fixed) parent node.  

Since the evaluation of the weight of every node configuration requires the calculation of the node's self energy, its interaction with its children and the weights of its children, the complexity, or number of elementary steps, required for assigning values to the spins of a tree of clustered spins, is linear in the number of edges (or nodes) in the tree. Specifically, it requires $2^{|\mathbf{s}_i|} \times 2^{|\mathbf{s}_j|}$ evaluations (where $|\mathbf{s}_i|$ is the number of spins in $\mathbf{s}_i$), for every pair of neighboring nodes $\mathbf{s}_i$ and $\mathbf{s}_j$:
\beq\label{eq:ct}
C_{\mathbf{t}}= \sum_{\langle \mathbf{s}_i, \mathbf{s}_j\rangle} 2^{|\mathbf{s}_i|}\times 2^{|\mathbf{s}_j|}
\approx 2^{2 \langle |\mathbf{s}| \rangle} (|\mathbf{t}|-1)
\eeq
where
\beq
\langle |\mathbf{s}| \rangle =\frac1{|\mathbf{t}|} \sum_i | \mathbf{s}_i |
\eeq
is the average number of spins per node and $|\mathbf{t}|$ is the size of the tree; i.e., the number of tree nodes (here, $|\mathbf{t}|-1$ is the number of edges).\footnote{In the above, we keep implicit the linear dependence of the complexity of the calculation on the degree (or number of neighbors) of the spins in the tree.}

\subsection{Locally optimal random trees of spin clusters~\label{sec:rand}}

Having computed the complexity of drawing a configuration from the Boltzmann distribution of an Ising model defined on a tree of clustered spins (or in the $T \to 0$ limit, the complexity of finding an optimal configuration on that tree), we turn to addressing the question of generating \emph{optimal} TOSCs for a given input problem---TOSCs from which it would be most advantageous in terms of overall equilibration time to draw Boltzmann configurations. On the one hand, the larger the subgraphs induced on the input problem graph are, the lower the number of `tree flips' required to thermalize the input graph. On the other hand, the complexity associated with thermalizing trees with large clusters may become prohibitively costly. 
To capture the needed balance between tree size and the complexity of drawing a thermal Boltzmann sample from it, we define the simple figure of merit $F$ --- the ratio of the complexity of drawing a thermal configuration from a tree $C_{\mathbf{t}}$ to the number of configurations on it:
\beq
F=\frac{C_{\mathbf{t}}}{2^{\sum_i | \mathbf{s}_i |}} = \frac{2^{2\langle |\mathbf{s}| \rangle} (|\mathbf{t}|-1)}{2^{|\mathbf{t}| \langle |\mathbf{s}| \rangle}} \,,
\eeq
which captures the relative ease with which one can draw a thermal configuration on the tree out of $2^{|\mathbf{t}| \langle |\mathbf{s}| \rangle}$ possible configurations. The smaller $F$ is, the more merit there is to its thermalization. 

We are now in a position to outline the first step of the TOSC algorithm in which a random TOSC is generated `on-the-fly' on the input graph, minimizing (locally) the figure of merit $F$. This first step is based on sequentially adding spins to a subgraph, re-evaluating $F$ at every stage, until a local minimum of $F$ is found, at which point the process is ended. 

\begin{figure}[ht] 
\subfigure[]{\includegraphics[width=0.98\columnwidth]{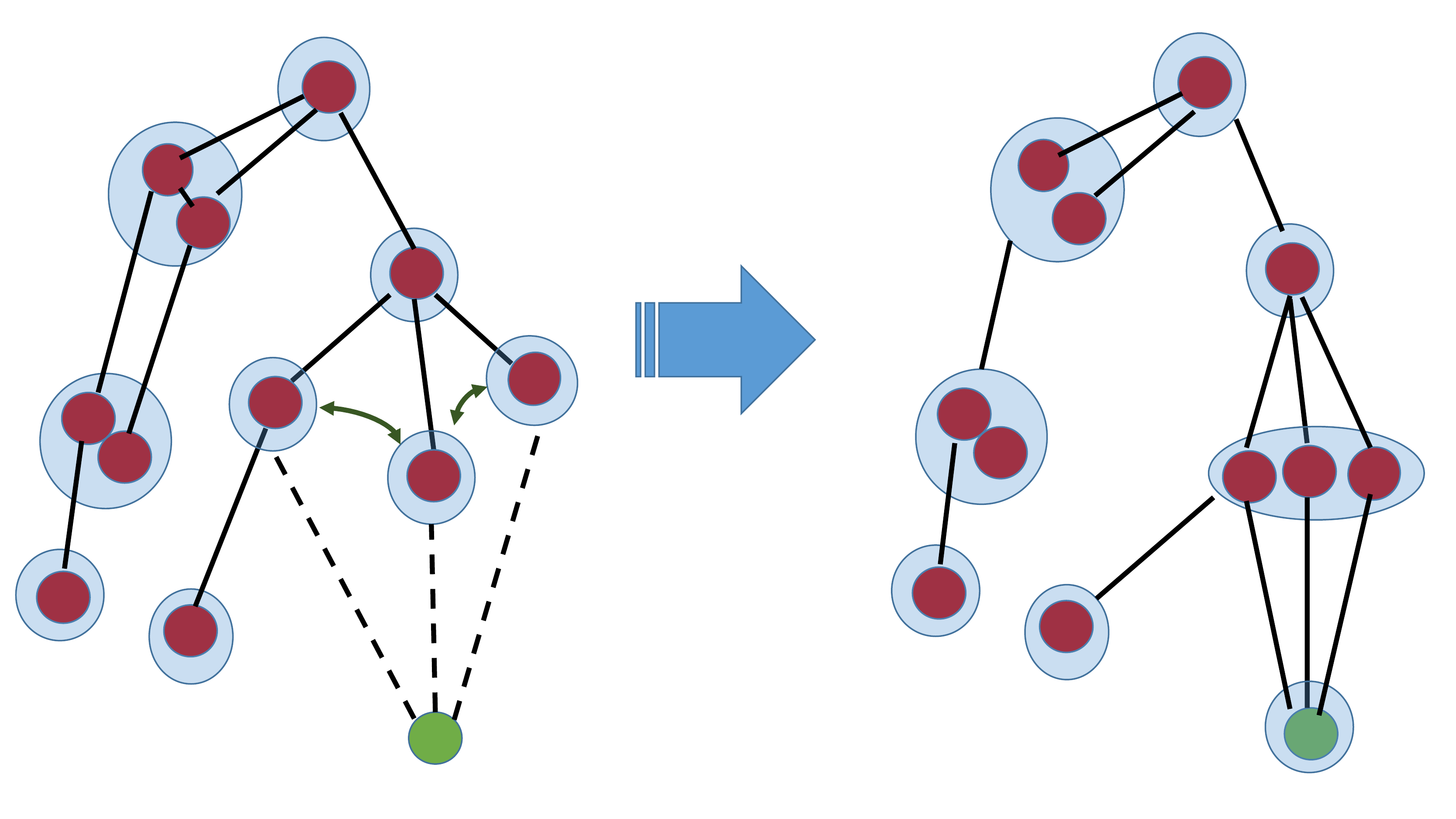}}
\subfigure[]{\includegraphics[width=0.98\columnwidth]{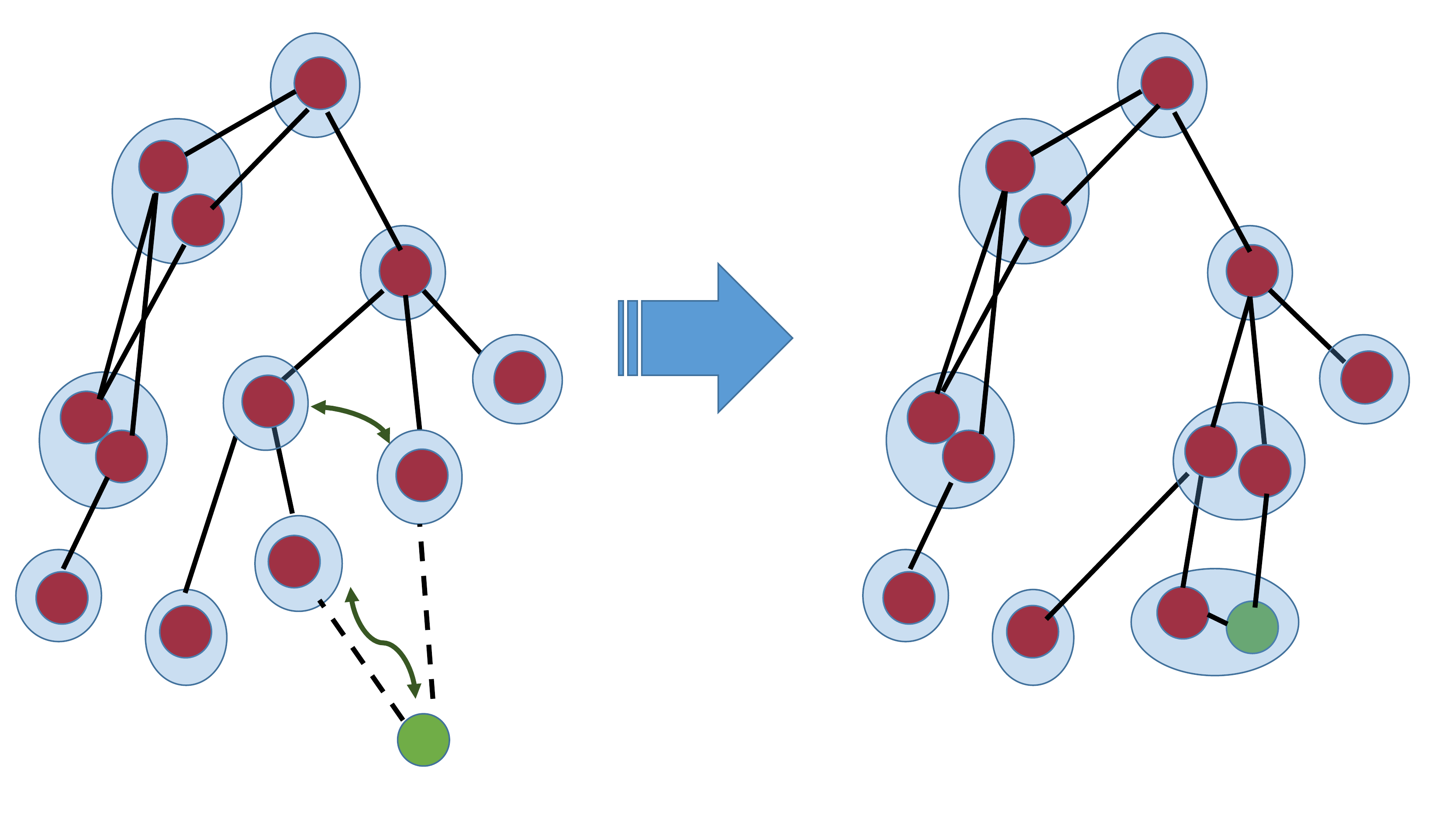}}
   \caption{{\bf Cycle contraction.} When considering the addition of a spin (green circle) to an existing tree that forms a cycle with the tree, the structure can retain its tree form if the cycle is contracted by joining together existing nodes. In both examples depicted above, the addition of the new spin closes a cycle, or cycles, and may therefore warrant a contraction, i.e., the unification of existing nodes. In the present algorithm, nodes that are equidistant from the root node are joined together. 
 After the cycle is contracted, the subgraph remains a tree of clustered spins (containing no cycles). }
   \label{fig:contraction}
\end{figure}

\begin{figure*}[ht] 
\subfigure[]{\includegraphics[width=0.66\columnwidth]{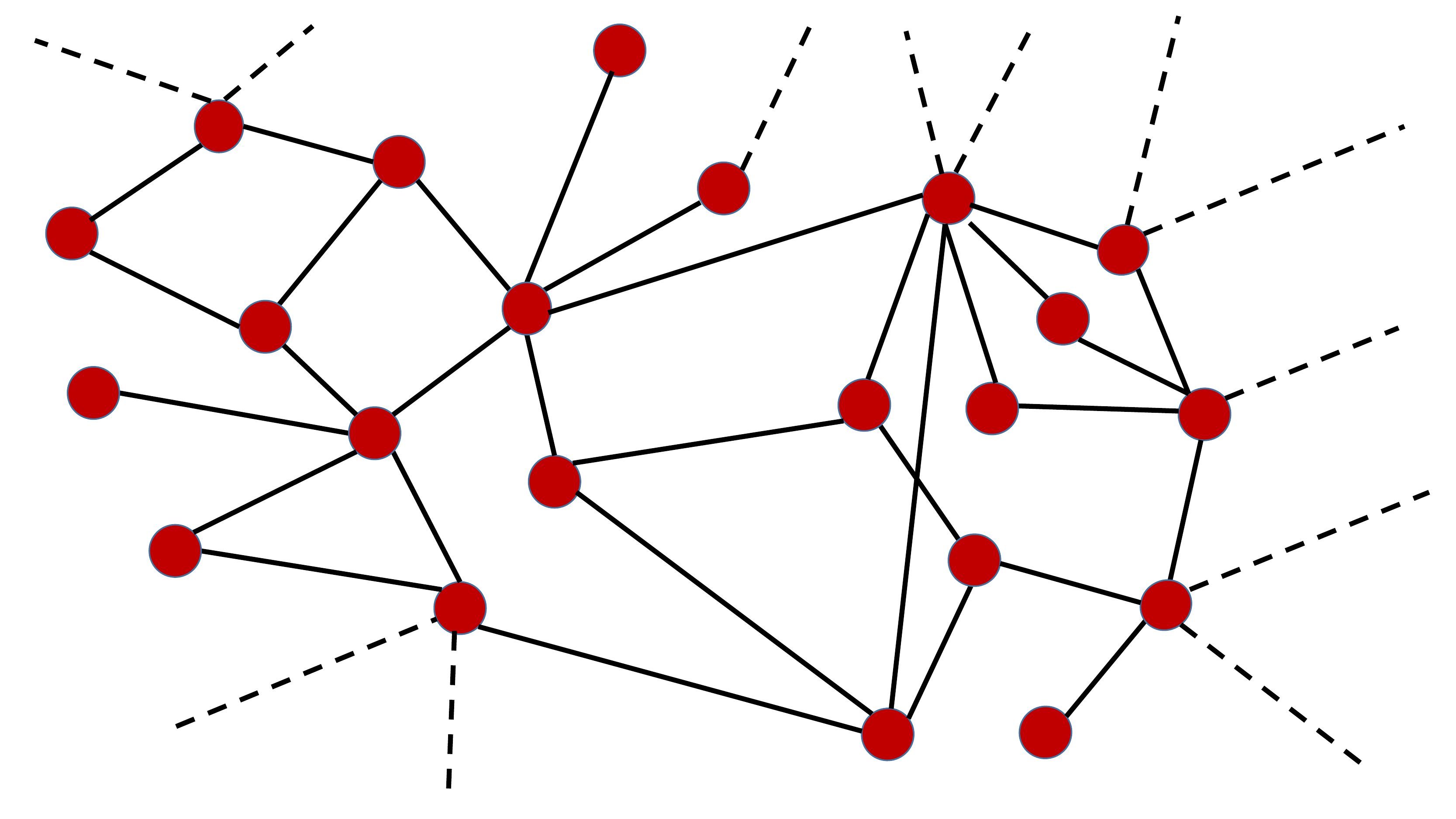}}
\subfigure[]{\includegraphics[width=0.66\columnwidth]{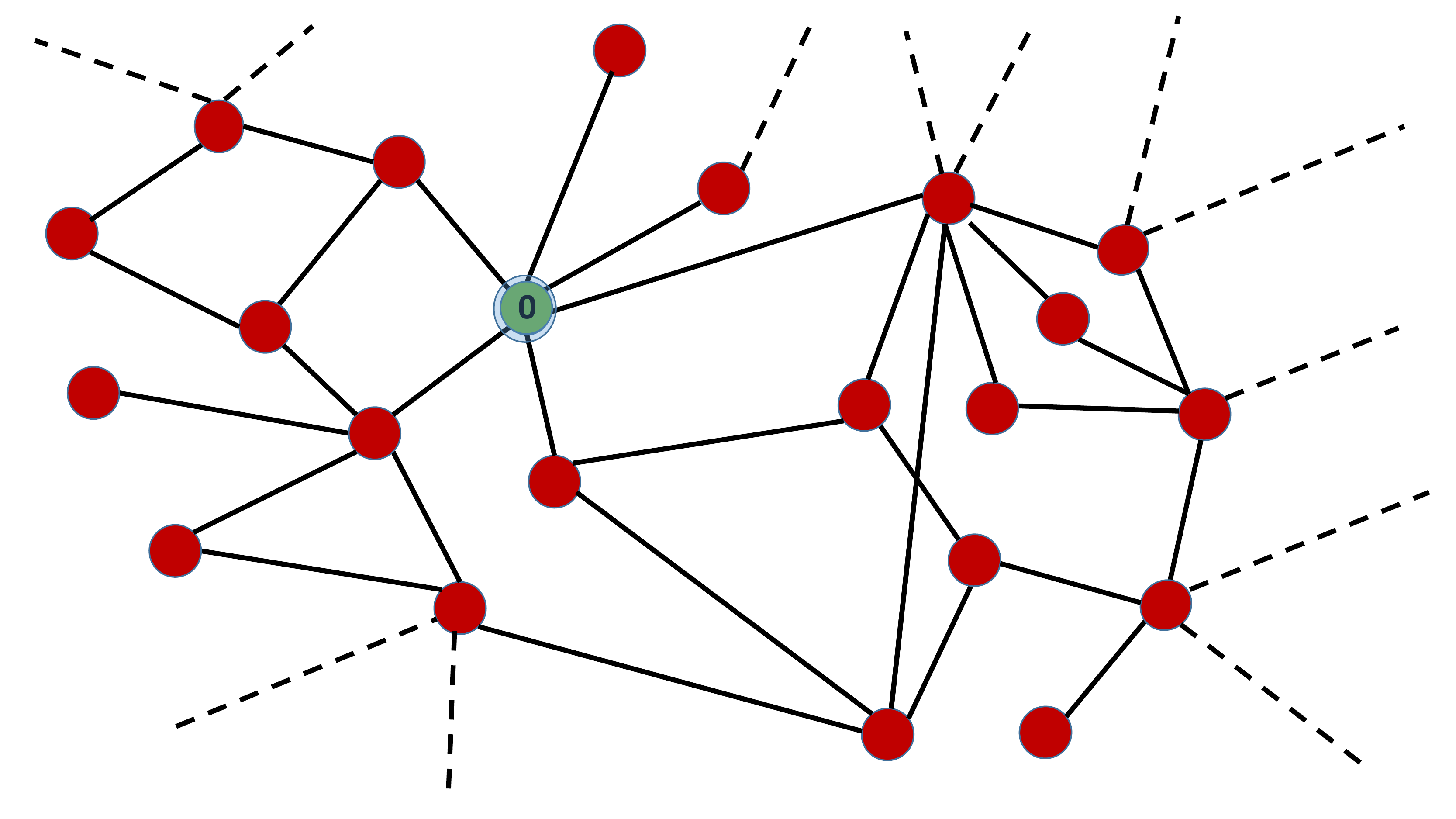}}
\subfigure[]{\includegraphics[width=0.66\columnwidth]{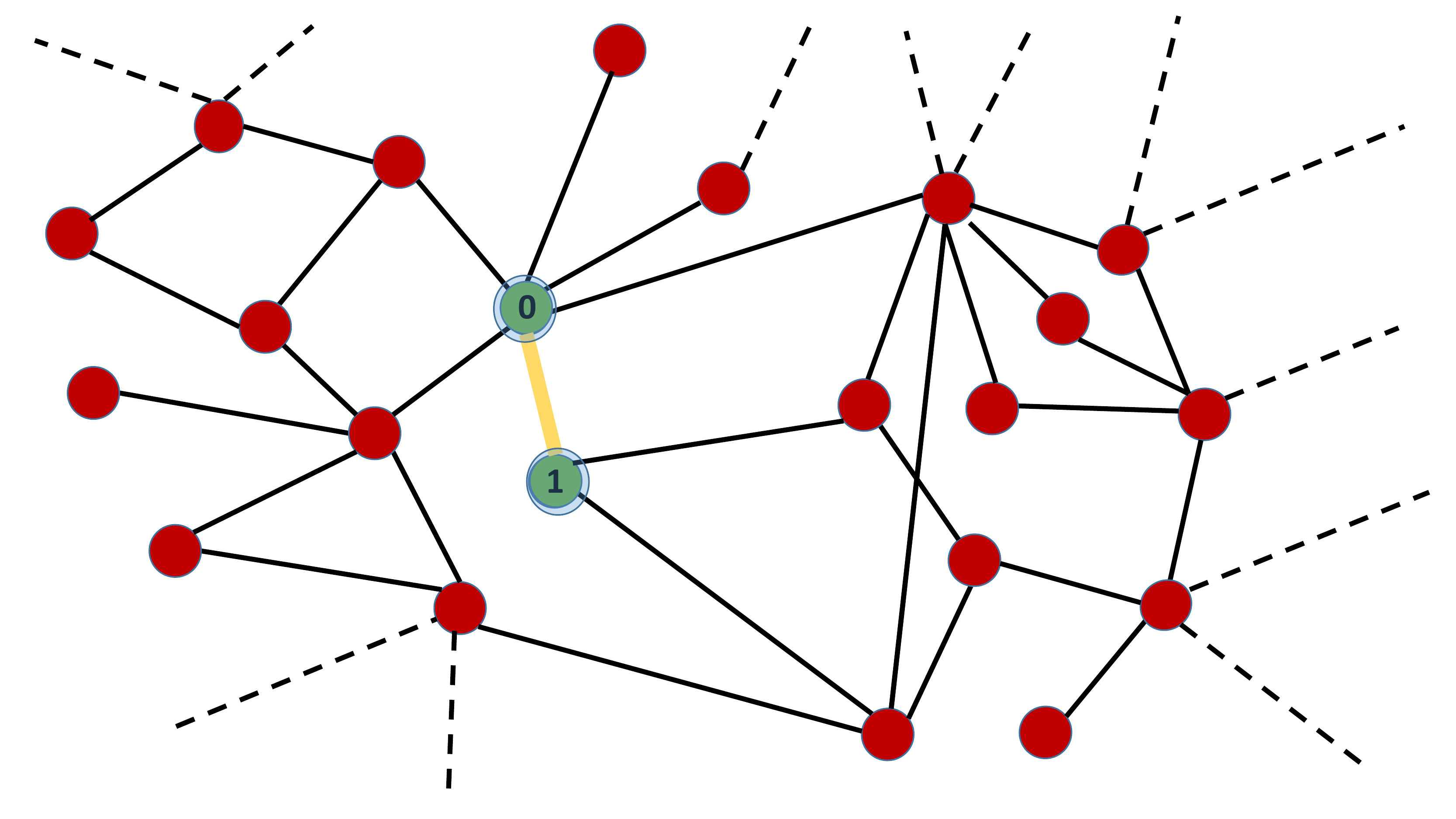}}
\subfigure[]{\includegraphics[width=0.66\columnwidth]{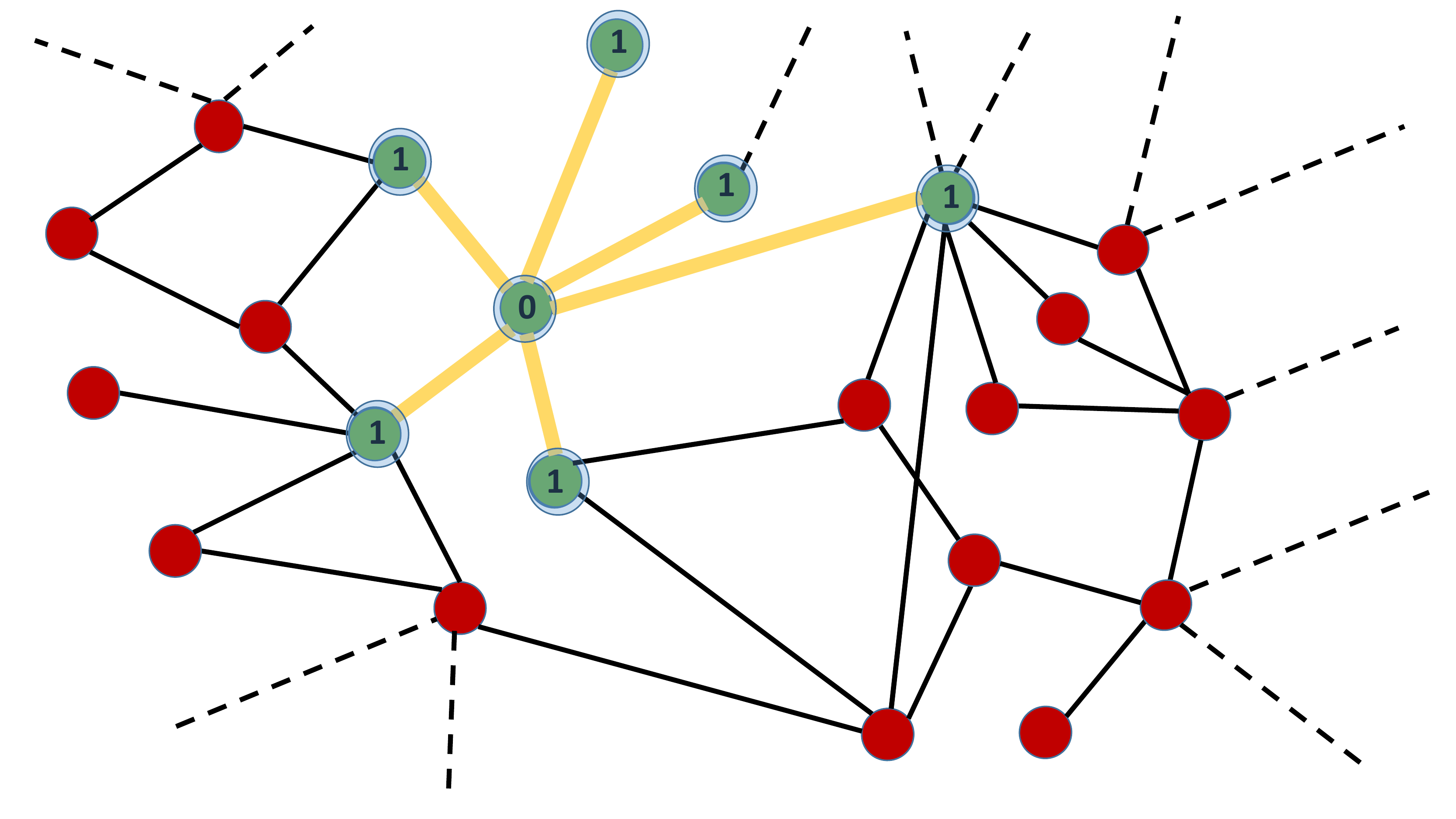}}
\subfigure[]{\includegraphics[width=0.66\columnwidth]{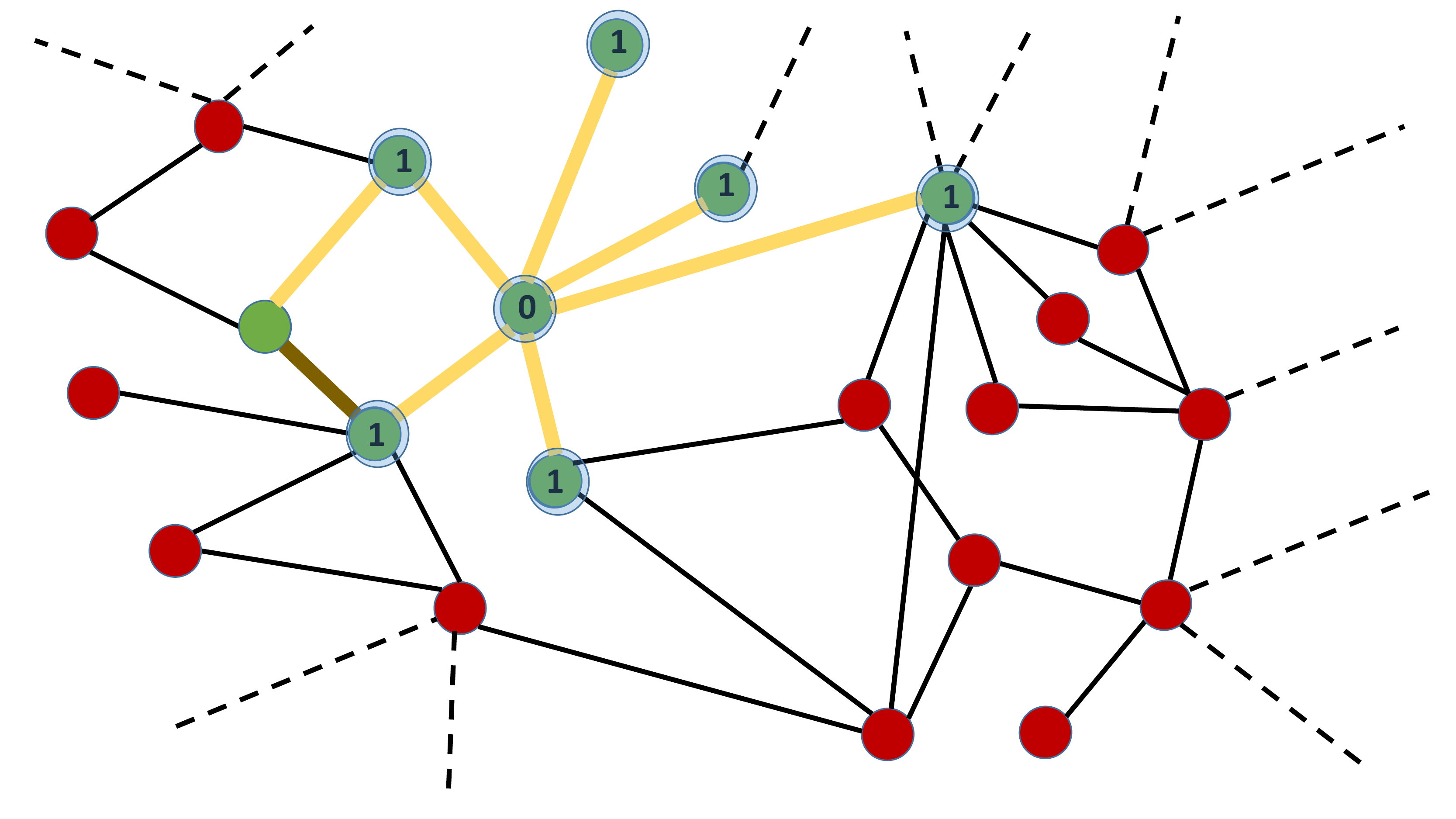}}
\subfigure[]{\includegraphics[width=0.66\columnwidth]{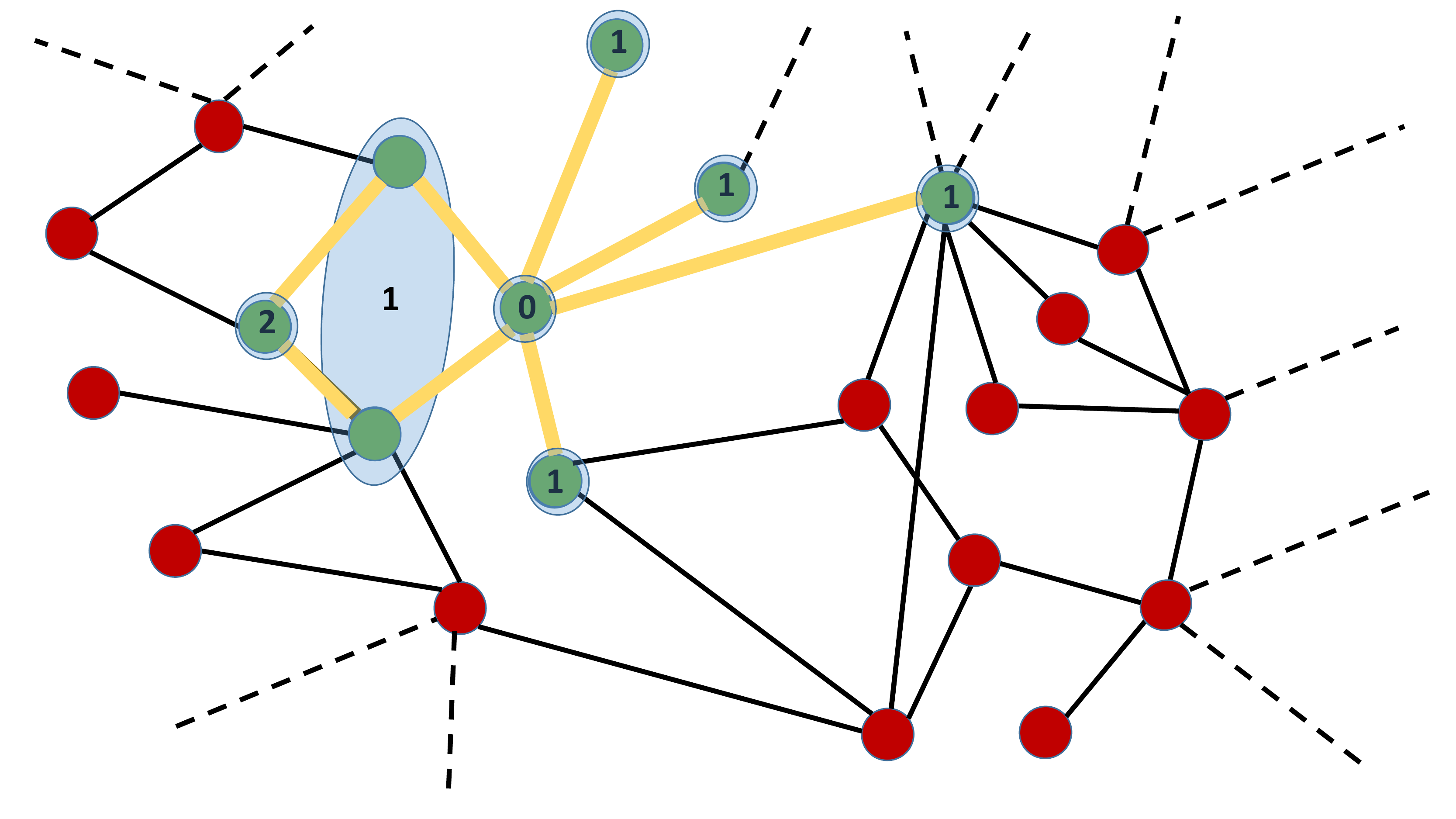}}
\subfigure[]{\includegraphics[width=0.66\columnwidth]{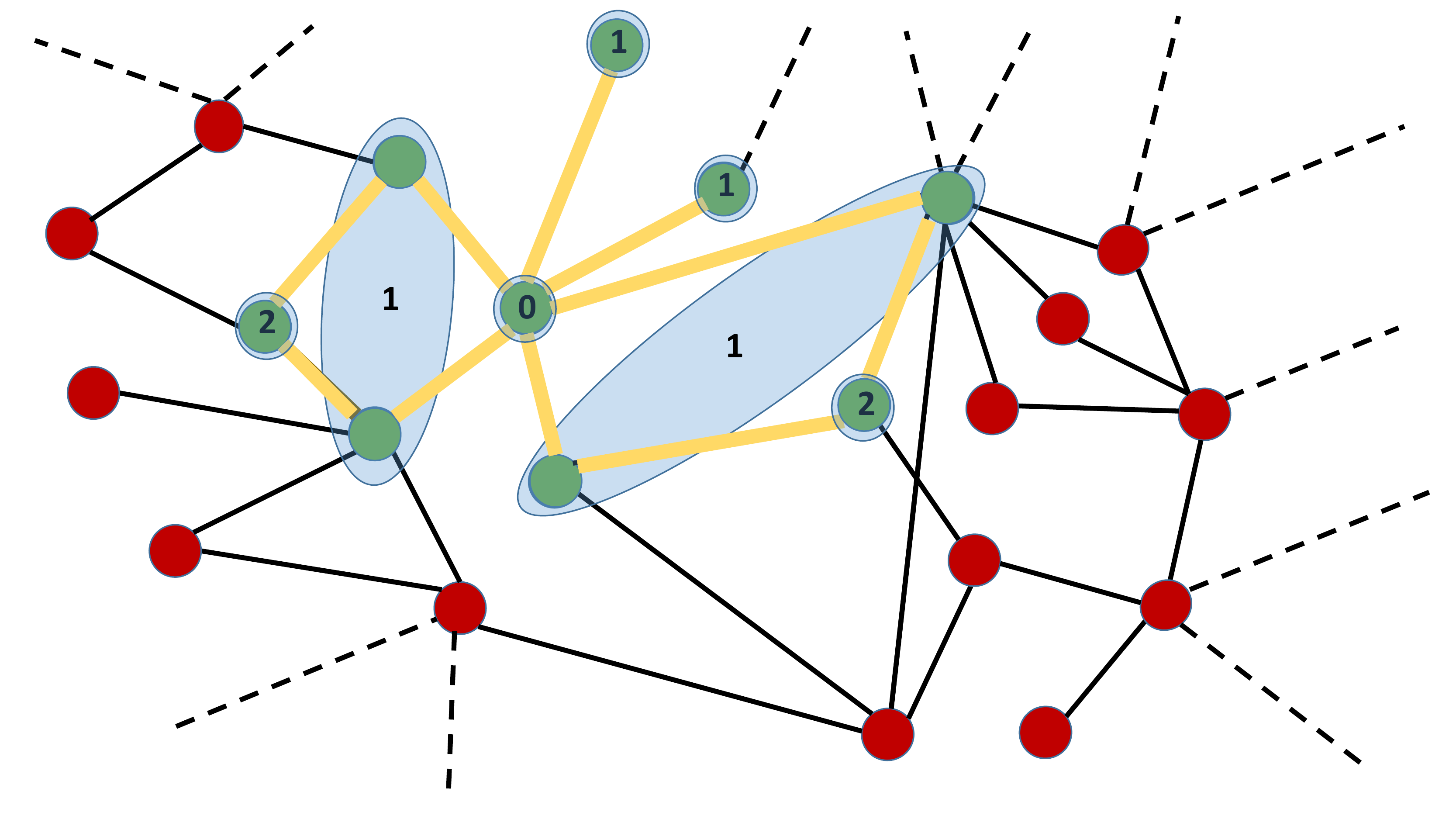}}
\subfigure[]{\includegraphics[width=0.66\columnwidth]{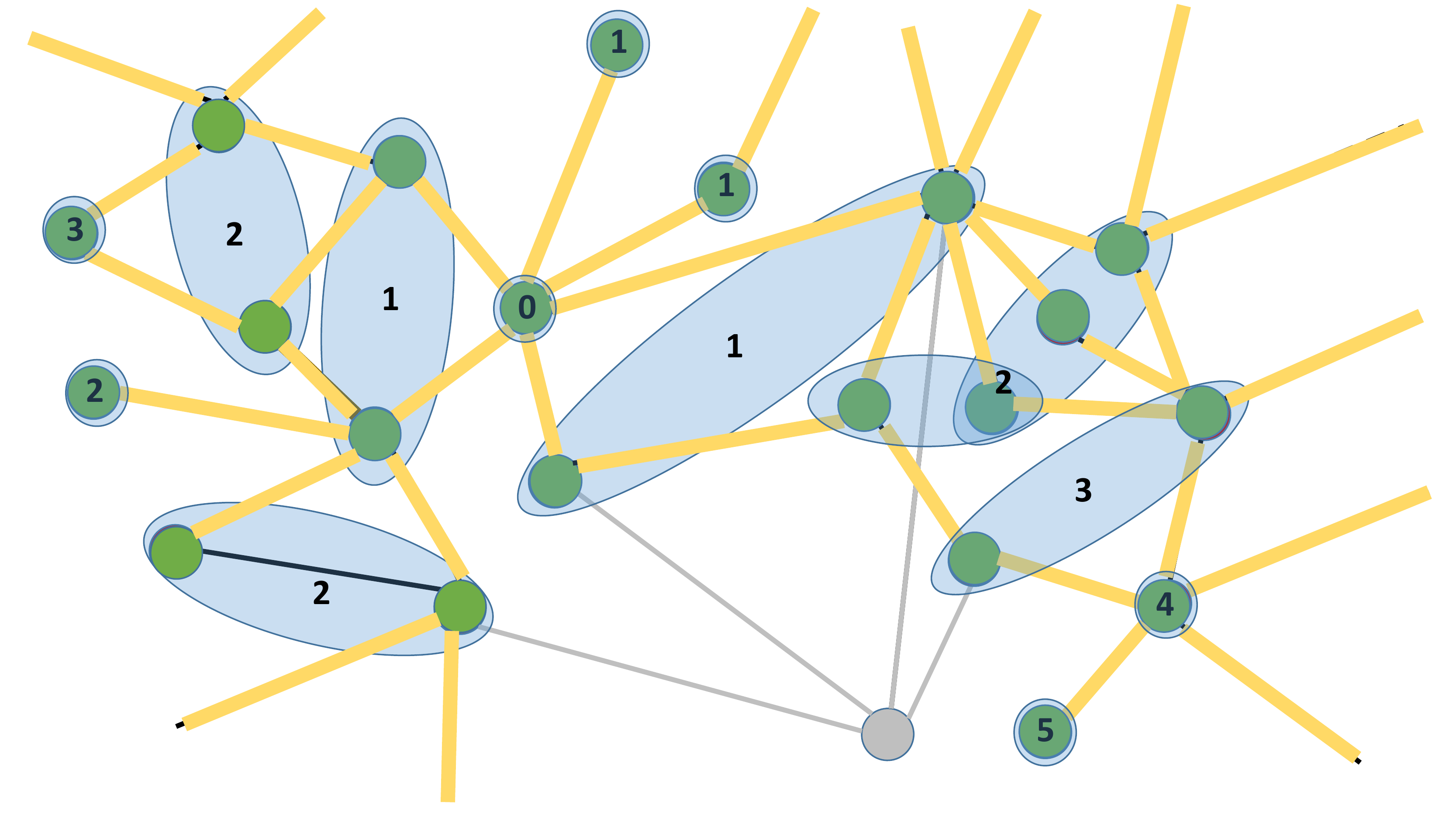}}
   \caption{{\bf A random construction of a tree of clustered spins.} (a) The input Ising problem graph. (b) A root node for the tree that is to be thermalized is chosen at random. (c) A random neighbor of the root is added to the growing tree. (d) All possible neighbors are added. (e)-(f) A `level-two' spin is chosen but can only be added on if two level-one spins `contract' to form a single cluster. (g) Another level-two spin is added. (h) The entire (visible) input graph, but one excluded spin, has become a tree of clustered spins. }
   \label{fig:randomGeneration}
\end{figure*}

The generation of a random TOSC on a given spin glass connectivity graph is carried out as follows. At first, a random spin that will function as a root node for the TOSC is selected. As a next step, its neighboring spins (or spins of distance one from the root) are added to the root, one by one, each forming a new neighboring tree node. 
The formation of a new node trivially lowers the value of $F$ and is thus always desired. If a spin that is to be added to the constructed tree is found to close a cycle, i.e., if it is connected to two or more different nodes on the tree, adding it as a new node as is, is prohibited. At this point, two alternatives are considered: i) The new spin is added to the TOSC by `contracting' the cycles formed due to its inclusion. This is done by unifying nodes in a manner that eliminates the cycles. Figure~\ref{fig:contraction} provides illustrations of cycle contraction, as the algorithm prescribes. ii) In the second alternative, the spin is not added to the TOSC being constructed and is taken out of the pool of addable spins. To determine which alternative is more beneficial, we evaluate $F$ for each scenario and pick the alternative with the smaller value. 
The process continues in a similar manner with the addition of spins that are neighbors of the neighbors of the root (equivalently, spins of distance two from the root) and so forth, until there are no additional spins to add.  Figure~\ref{fig:randomGeneration} illustrates the formation of a random TOSC embedded in a given input problem. 

The full algorithm thus consists of two steps: the first in which a random TOSC is generated by optimizing the benefit of thermalizing large subgraphs, and the second which actually draws a thermal sample from the tree. 

\subsection{Parallel tempering with TOSC\label{sec:PT}}

A way to further speed up the TOSC algorithm is to embed it in a parallel tempering (PT) scheme. 
In PT, one considers $N_T$ independent copies of the system running in parallel at different
temperatures, $T_1<T_2<\ldots< T_{N_T}$.\footnote{Here, we choose a temperature grid of the PT simulations consisting of $N_T=30$ temperatures. Temperatures with
indices $i=13,14,\ldots, N_T$ were evenly distributed in the range
$0.21 \leq T_i \leq T_\text{max}=1.632$, while lower temperatures in
the range $T_\mathrm{min}=0.045 \leq T_i \leq 0.2$ (indices
$i=1,2,\ldots,12$)~\cite{scirep15:Martin-Mayor_Hen}.}
Copies with neighboring temperatures regularly attempt to swap their
temperatures with probabilities that satisfy detailed balance~\cite{sokal:97}. In this
way, each copy performs a temperature random-walk. At high temperatures, free-energy
barriers are easily overcome, allowing for a global exploration of
configuration space. At lower temperatures on the other hand, the local
minima are explored in more detail.  
We will consider a variant of the algorithm where a single TOSC is generated for all replicas at the beginning of each `temperature sweep,' and during the sweep random samples are drawn on the tree for each copy independently, based on its temperature. 

\section{Algorithm benchmarking\label{sec:results}}

We now present the results of several benchmarking tests performed on the proposed algorithm. 
We compare the typical runtimes of three parallel tempering-based optimization algorithms: (i) single spin-flip Metropolis algorithm (SSF), (ii) parallel tempering with randomly generated trees of single spins (TSS), and (iii) the TOSC algorithm. 

Here, the typical runtimes for any set of parameters is defined as the median time to reach a minimizing configuration (i.e., a ground state), or thermal equilibrium,  over 100 randomly generated instances with the specified parameters. All the algorithms were run on one core of a $3.5$ GHz 6-Core Intel Xeon E5 processor. 

\subsection{Optimization of trees of fixed-size spin clusters}

\begin{figure}[ht] 
\includegraphics[width=0.99\columnwidth]{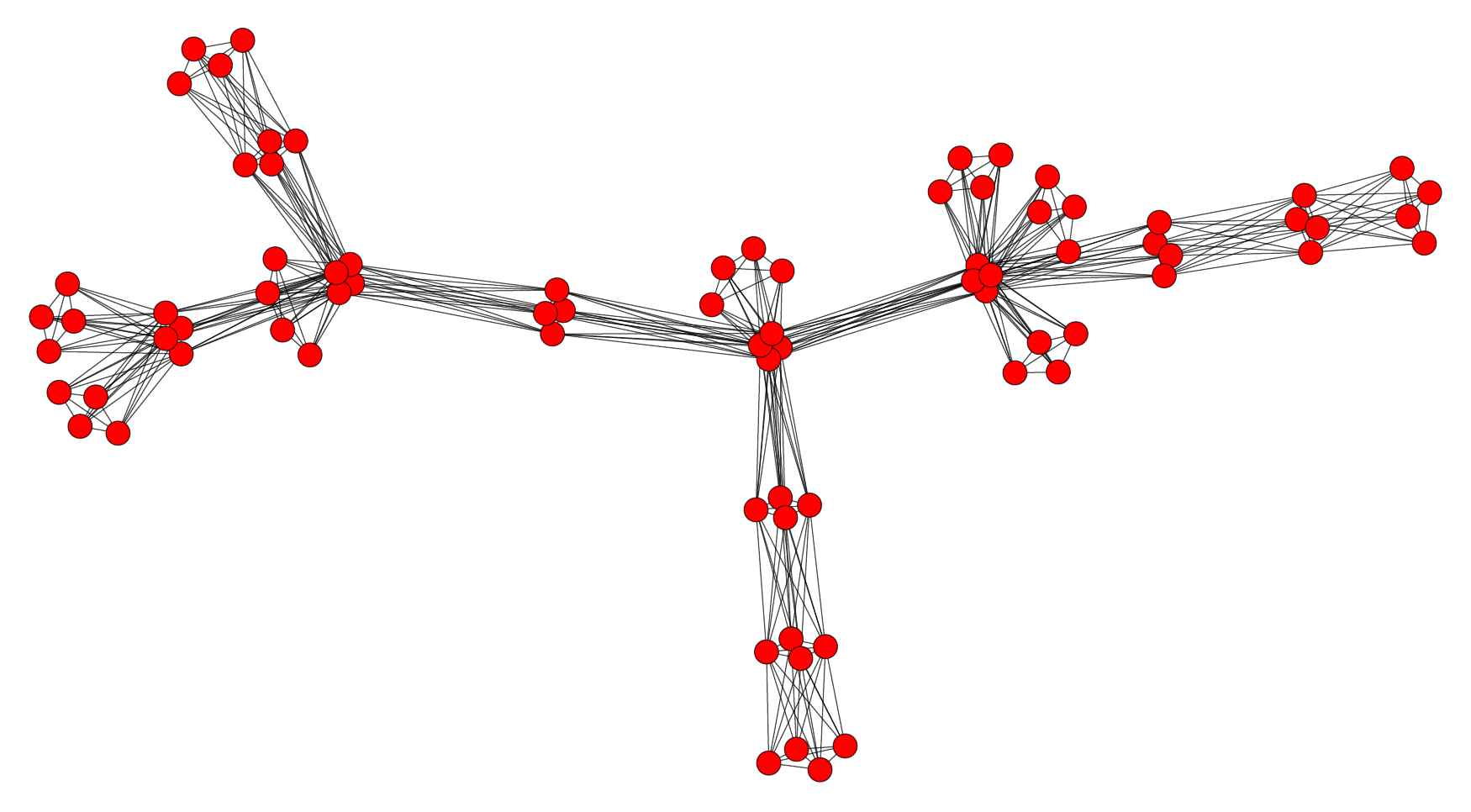}
   \caption{{\bf A tree of fixed-size spin clusters.} The connectivity graph is a randomly generated tree with $20$ four-spin nodes. Edges exist between all the spins within a given node and between spins of neighboring nodes.}
   \label{fig:randomTocs}
\end{figure}

The first class of random Ising problems we test is one where the underlying connectivities of the input problems themselves are randomly generated trees of spin clusters of fixed-size $C$, wherein all the spins within a cluster are connected by edges, and edges also exist between all spin pairs that belong to neighboring clusters. An example of a random $20$-node input problem  with $C=4$ spins per cluster (and $80$ spins in total) is given in Fig.~\ref{fig:randomTocs}. The coupling strengths $J_{ij}$ on each edge are chosen to be $\pm1$ with equal probability, and the external fields $h_i$ are set to zero. 

\begin{figure*}[ht] 
\subfigure[]{\includegraphics[width=0.51\columnwidth]{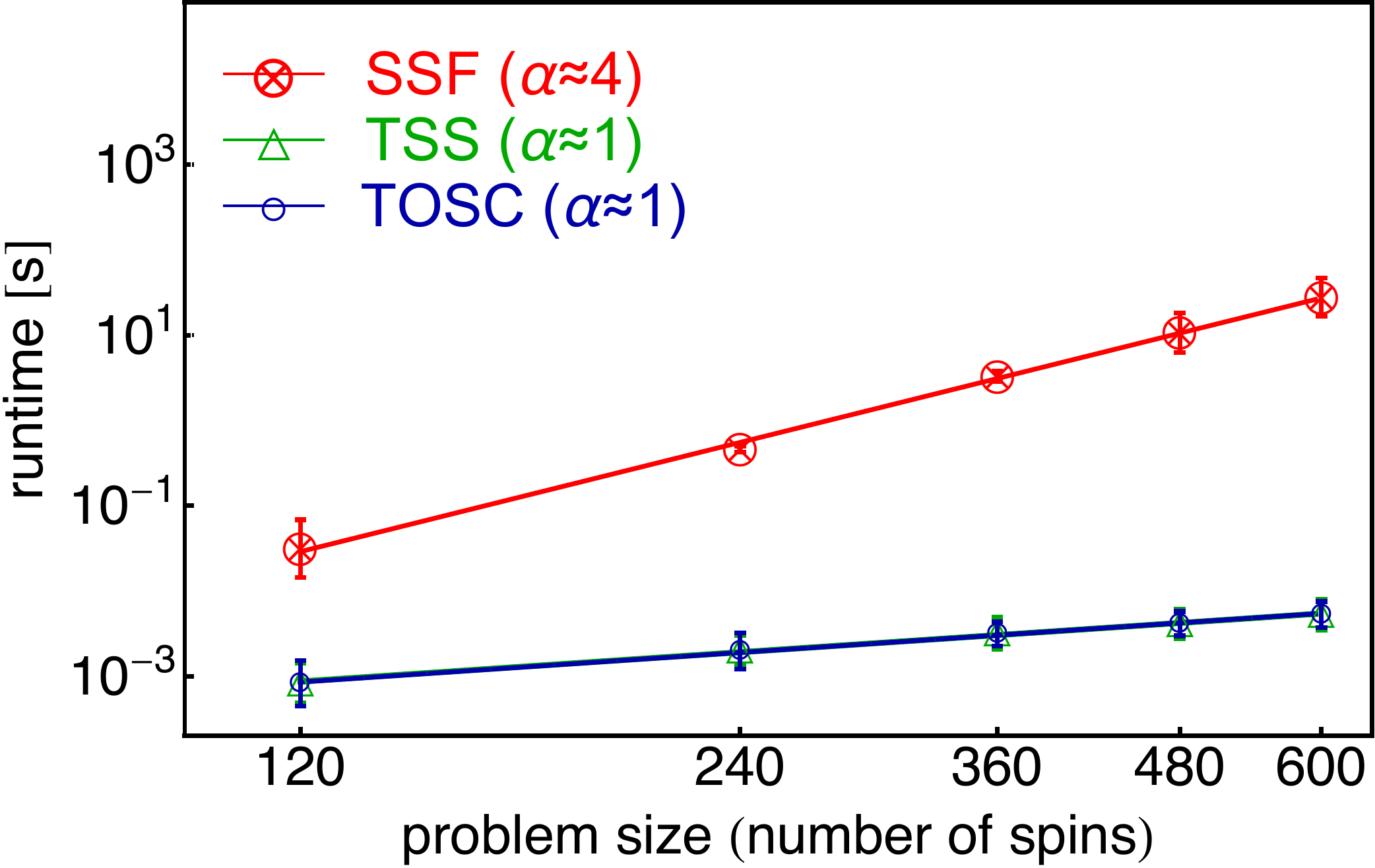}}
\subfigure[]{\includegraphics[width=0.51\columnwidth]{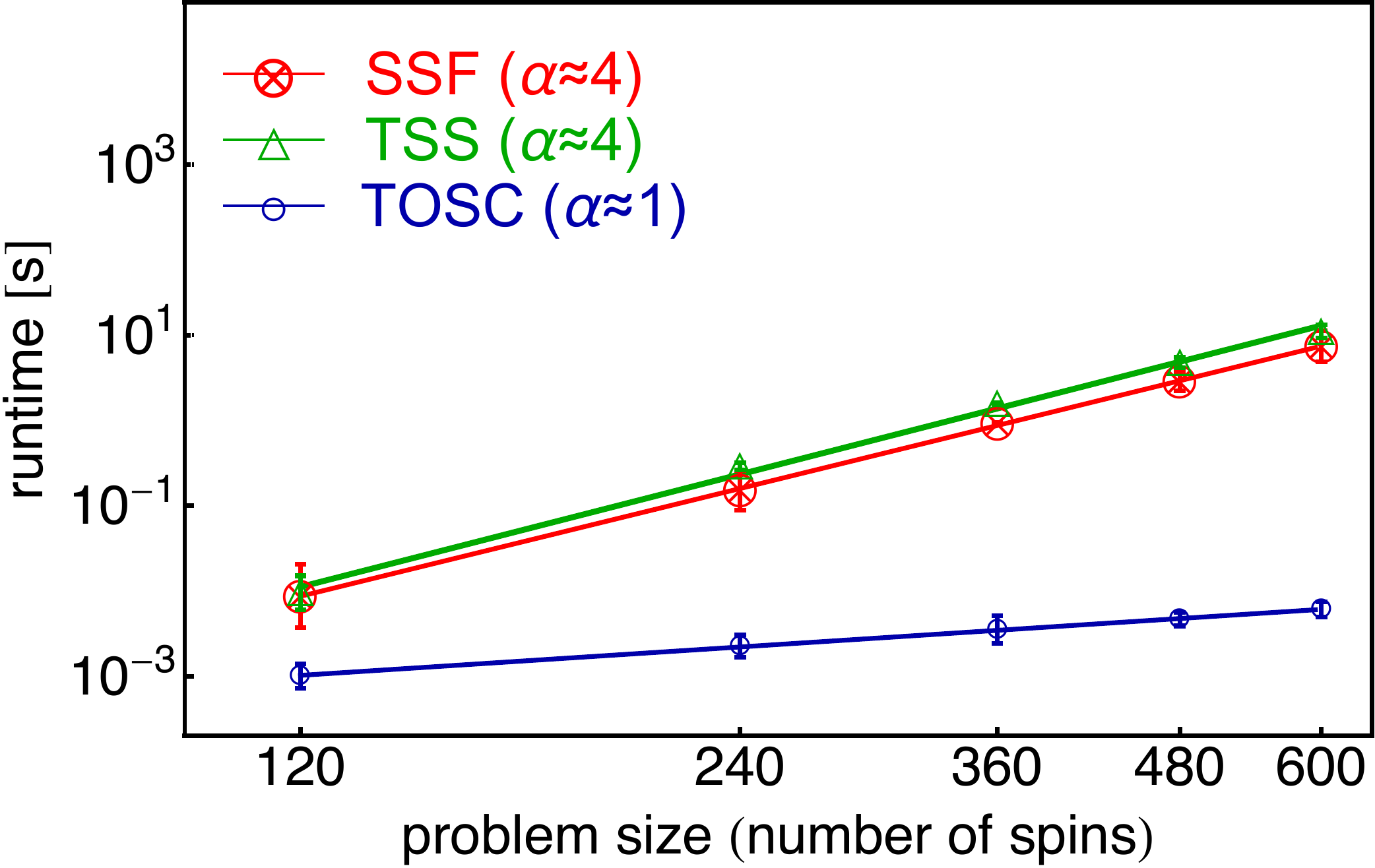}}
\subfigure[]{\includegraphics[width=0.51\columnwidth]{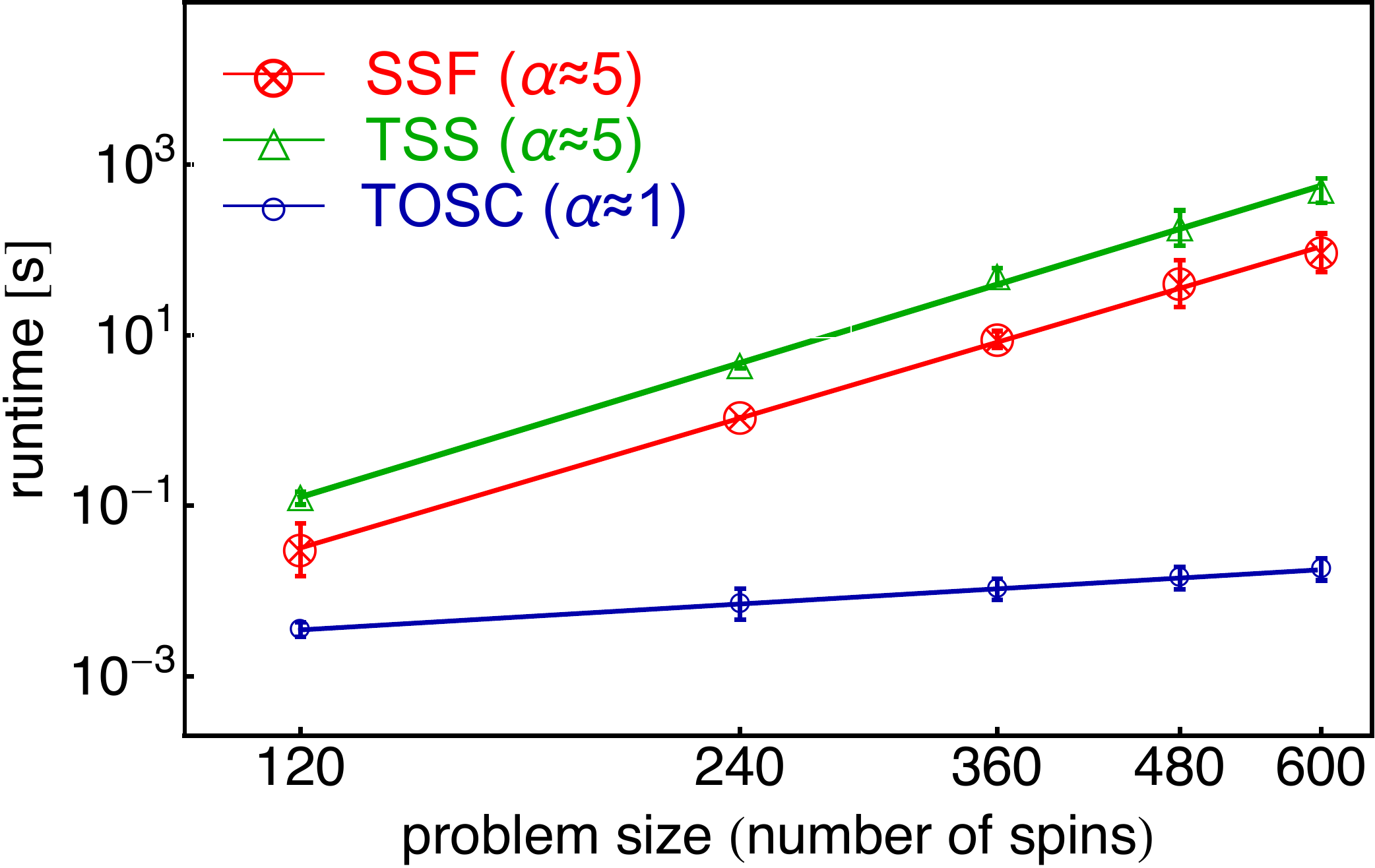}}
\subfigure[]{\includegraphics[width=0.51\columnwidth]{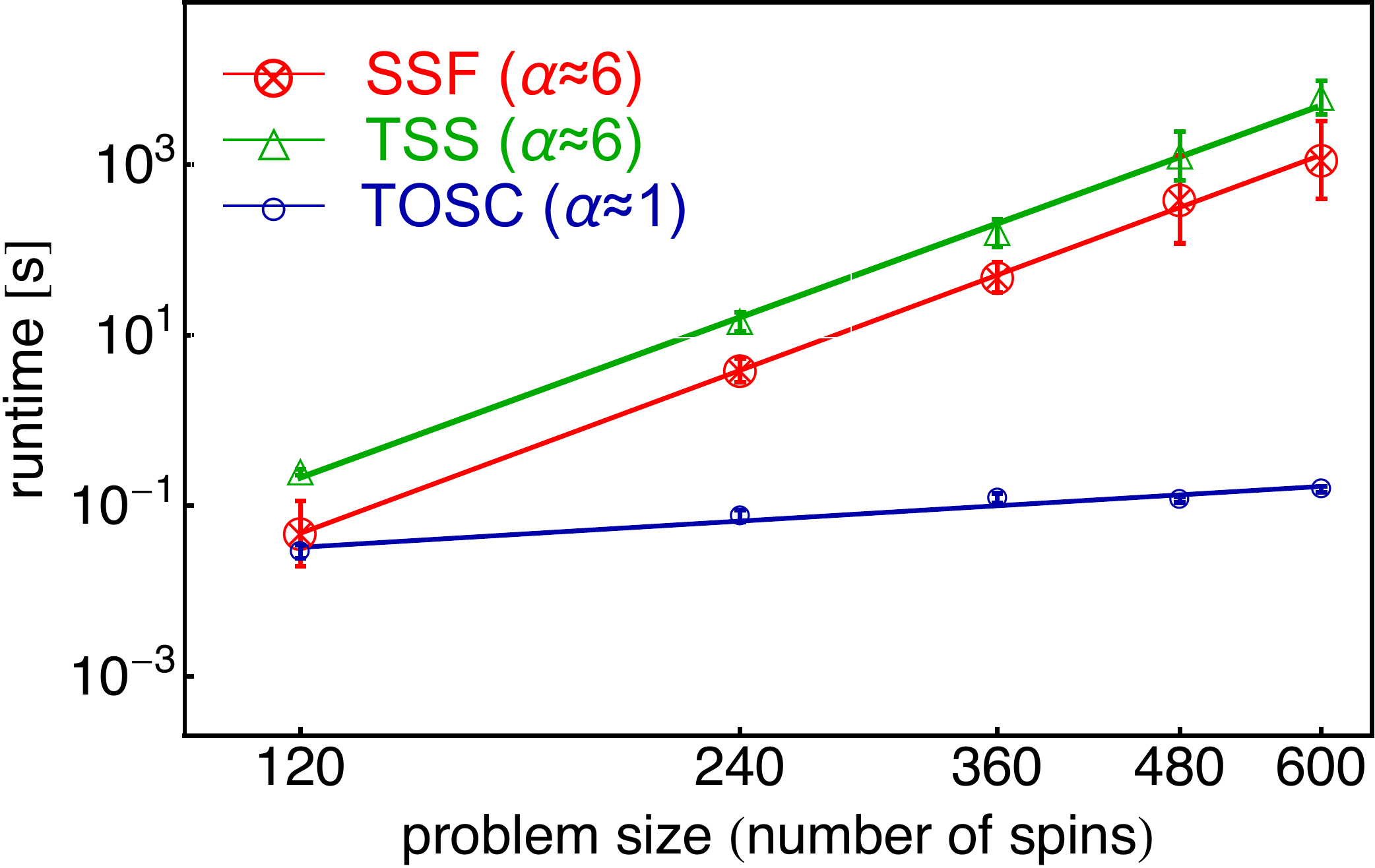}}
   \caption{{\bf Typical optimization runtimes of single spin-flip (SSF) parallel tempering, random trees of single spins (TSS) and optimized random trees of spin clusters (TOSC) on random instances of trees of fixed-size clusters.} Runtimes are shown on a log-log scale as a function of problem size. From left to right, the cluster sizes considered are $C=1, 2, 4$ and $6$. While all three algorithms exhibit power-law scaling, the SSF and TSS algorithms scale with powers that grow with $C,$ whereas the TOSC algorithm runtime scales linearly with problem size regardless of cluster size.}
   \label{fig:resRandomTrees}
\end{figure*}

For this type of problem, an a priori knowledge of the structure of the input problem allows for the devising of an optimization algorithm that takes full advantage of the underlying tree structure of the problem. In this case, the runtime scales linearly with the number of clusters for any fixed $C$. 
In Fig.~\ref{fig:resRandomTrees}, we show the typical runtimes of the SSF, TSS and TOSC algorithms as a function of problem size for different values of cluster size $C$.
As is immediately evident, the typical runtimes of all algorithms scale polynomially with problem size. However, for both SSF and TSS the power (the slope) grows rapidly with increasing cluster size $C$, whereas the TOSC runtime scales linearly with problem size irrespective of $C$, similar to the ideal algorithm. For $600$-spin problems, this amounts to speed-ups of up to four orders of magnitude of TOSC as compared to the other algorithms.

\begin{figure}[ht] 
\subfigure[]{\includegraphics[width=0.8\columnwidth]{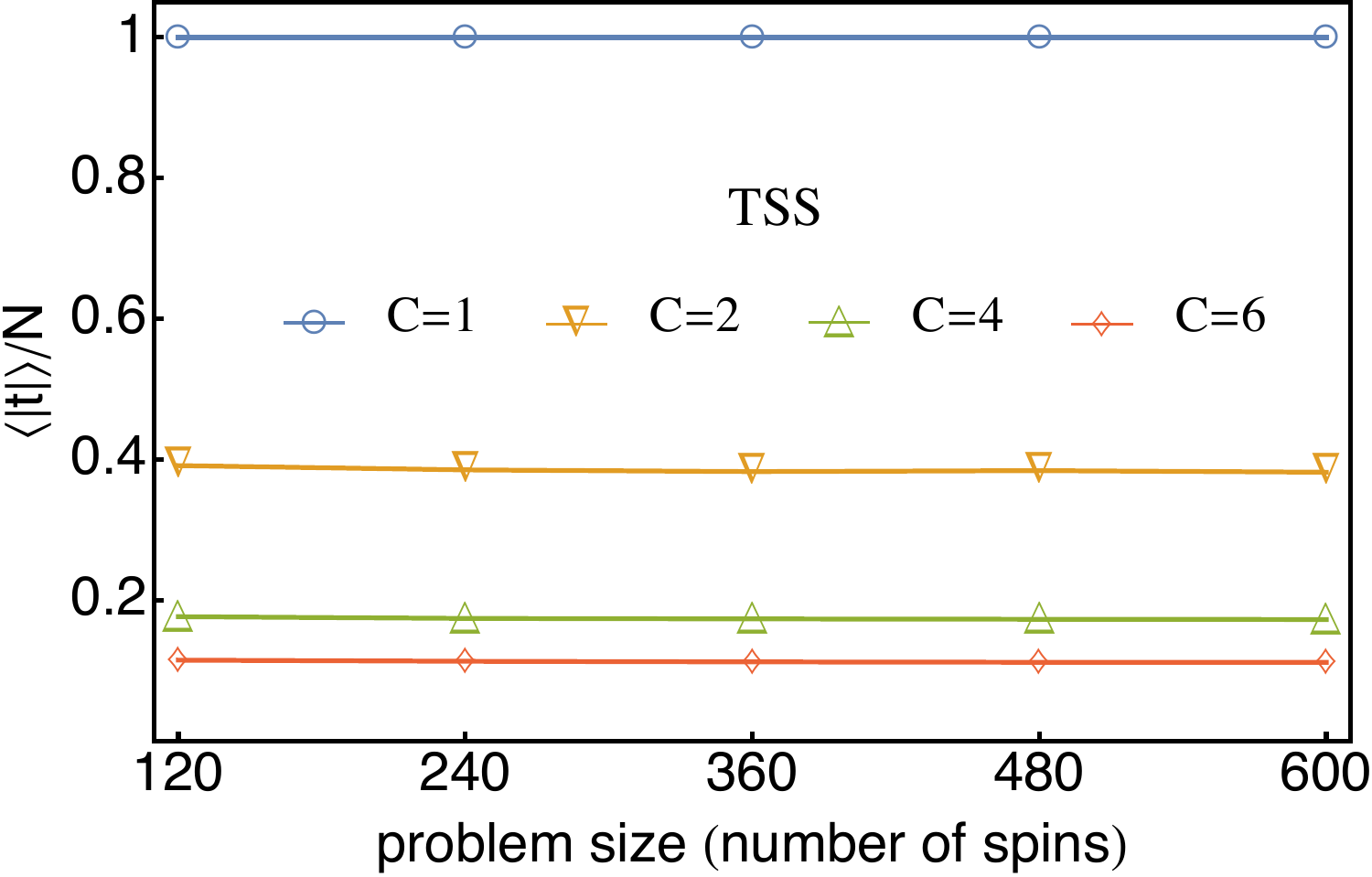}}
\subfigure[]{\includegraphics[width=0.8\columnwidth]{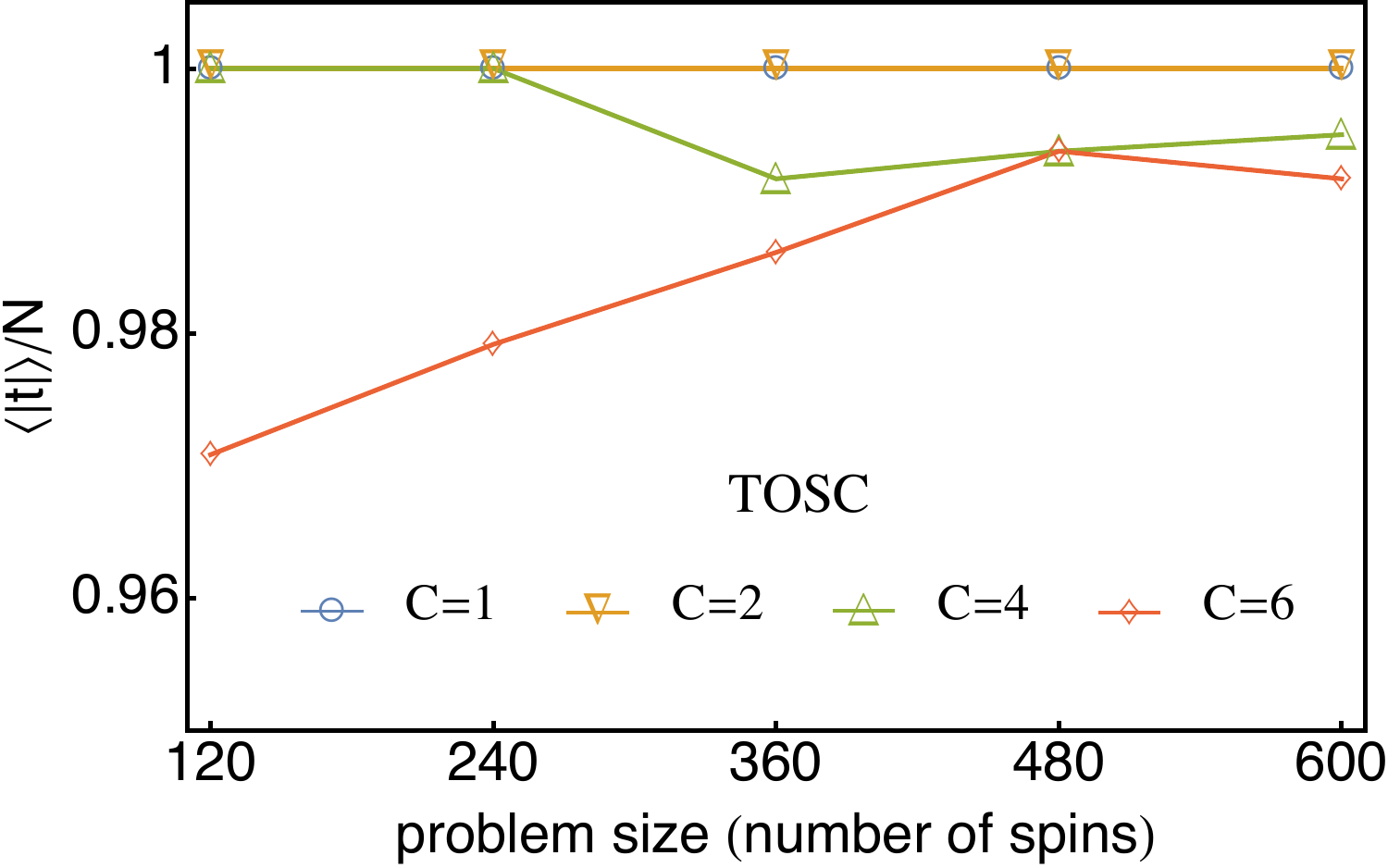}}
   \caption{{\bf Average ratio of tree size $\langle |\mathbf{t}|\rangle$ to problem size $N$ of TSS and TOSC on random instances of trees of fixed-size clusters.} While trees of single spins cover less and less of the input graph with growing cluster size $C$, the TOSC tree sizes tend to cover the entire input graph in the large problem size limit.}
   \label{fig:treeStatRandomTrees}
\end{figure}

It is also instructive to look at the average ratio of tree size $\langle |\mathbf{t}|\rangle$ to problem size $N$ of the trees generated by TSS and TOSC. 
This is shown in Fig.~\ref{fig:treeStatRandomTrees}. While the TSS tree covers less and less of the input problem with growing cluster size [Fig.~\ref{fig:treeStatRandomTrees}(a)], the TOSC trees tend to cover the entire input graph in the limits of large problem sizes [Fig.~\ref{fig:treeStatRandomTrees}(b)]. 

\subsection{Thermalization of trees of fixed-size spin clusters}

To test the performance of the TOSC algorithm as a thermalizer rather than an optimizer, that is, to ascertain its ability to draw configurations from the Boltzmann distributions of given Hamiltonians at given temperatures, we next compare the typical runtimes of SSF, TSS and TOSC to equilibrate trees of fixed-size spin clusters. To that aim, we utilize the concept of PT mixing time $\tau$~\cite{fernandez:09b,janus:10,fernandez:13,scirep15:Martin-Mayor_Hen,marshall:16} using it as the figure of merit for thermalization. The mixing time of a PT simulation may be thought of as the average time it takes a PT replica to fully traverse the temperature mesh, indicating equilibration of the simulation. In PT, copies
with neighboring temperatures regularly attempt to swap their
temperatures with probabilities that satisfy detailed balance~\cite{sokal:97}. In this
way, each copy performs a temperature random-walk. At high temperatures, free-energy
barriers are easily overcome, allowing for a global exploration of
configuration space. At lower temperatures on the other hand, the local
minima are explored in more detail.  A `healthy' PT simulation
requires an unimpeded temperature flow and so the total length of the
simulation should  be longer than the temperature mixing
time~\cite{fernandez:09b,janus:10}. 

\begin{figure*}[ht] 
\subfigure[]{\includegraphics[width=0.661\columnwidth]{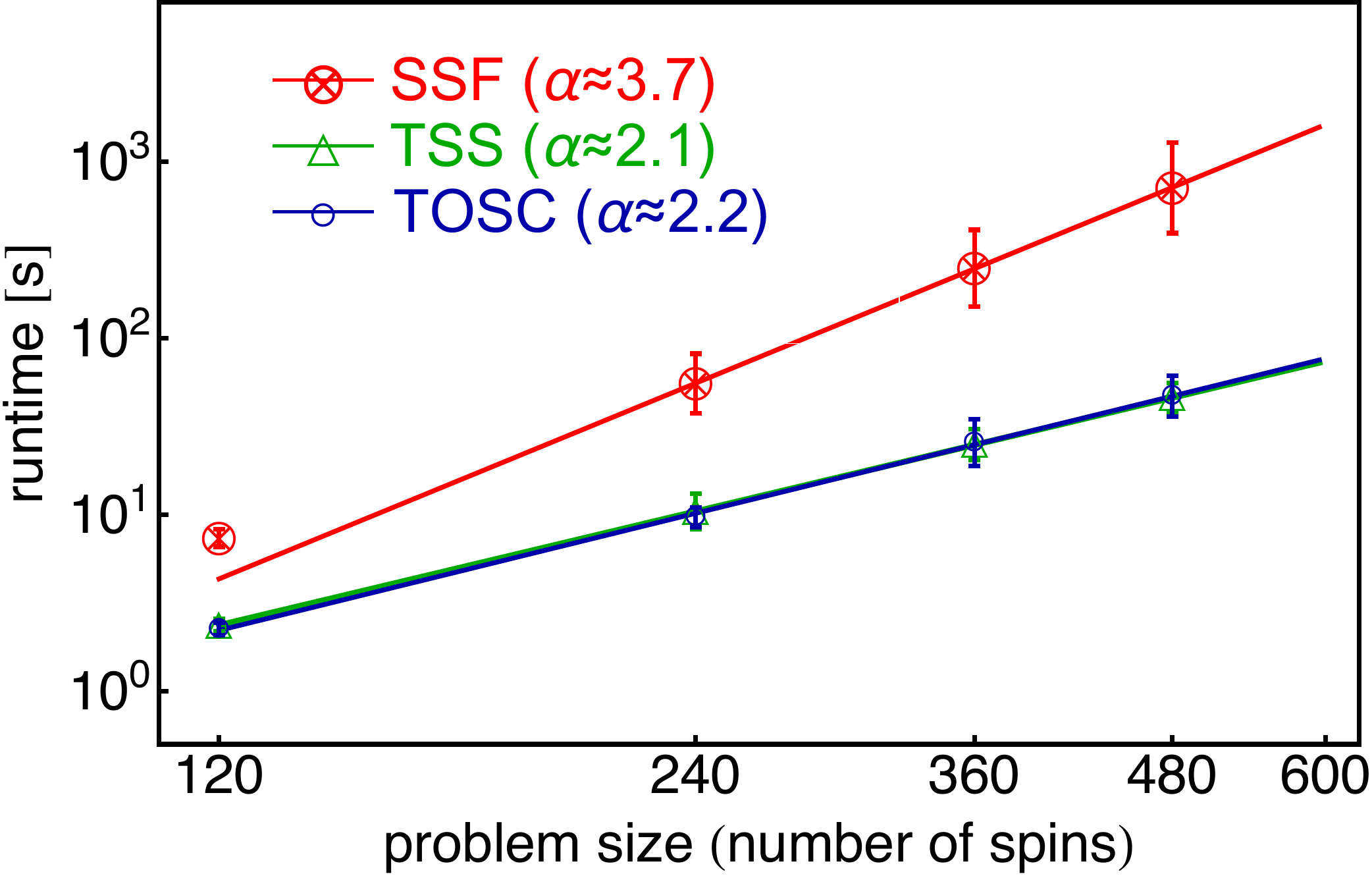}}
\subfigure[]{\includegraphics[width=0.661\columnwidth]{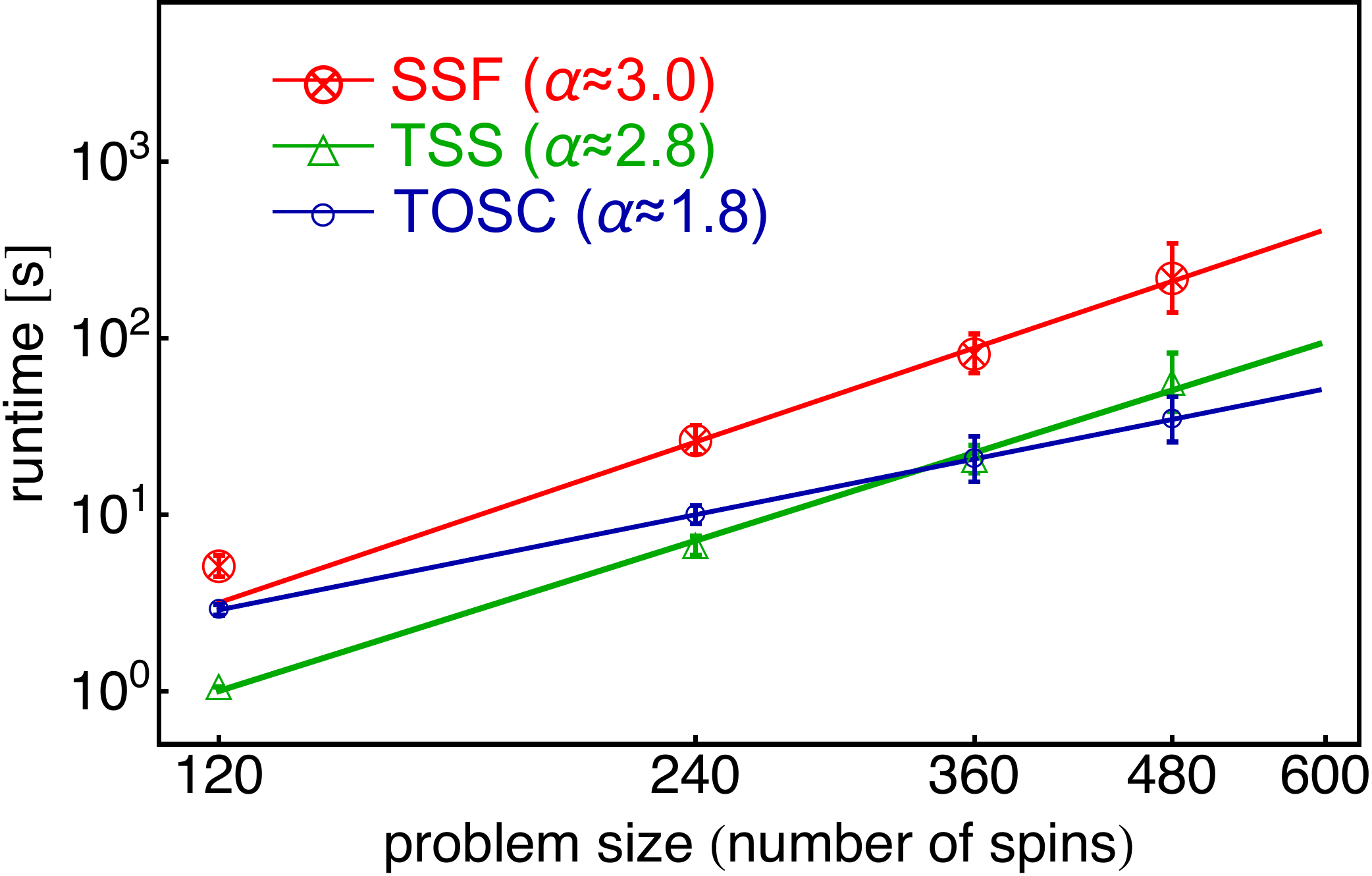}}
\subfigure[]{\includegraphics[width=0.661\columnwidth]{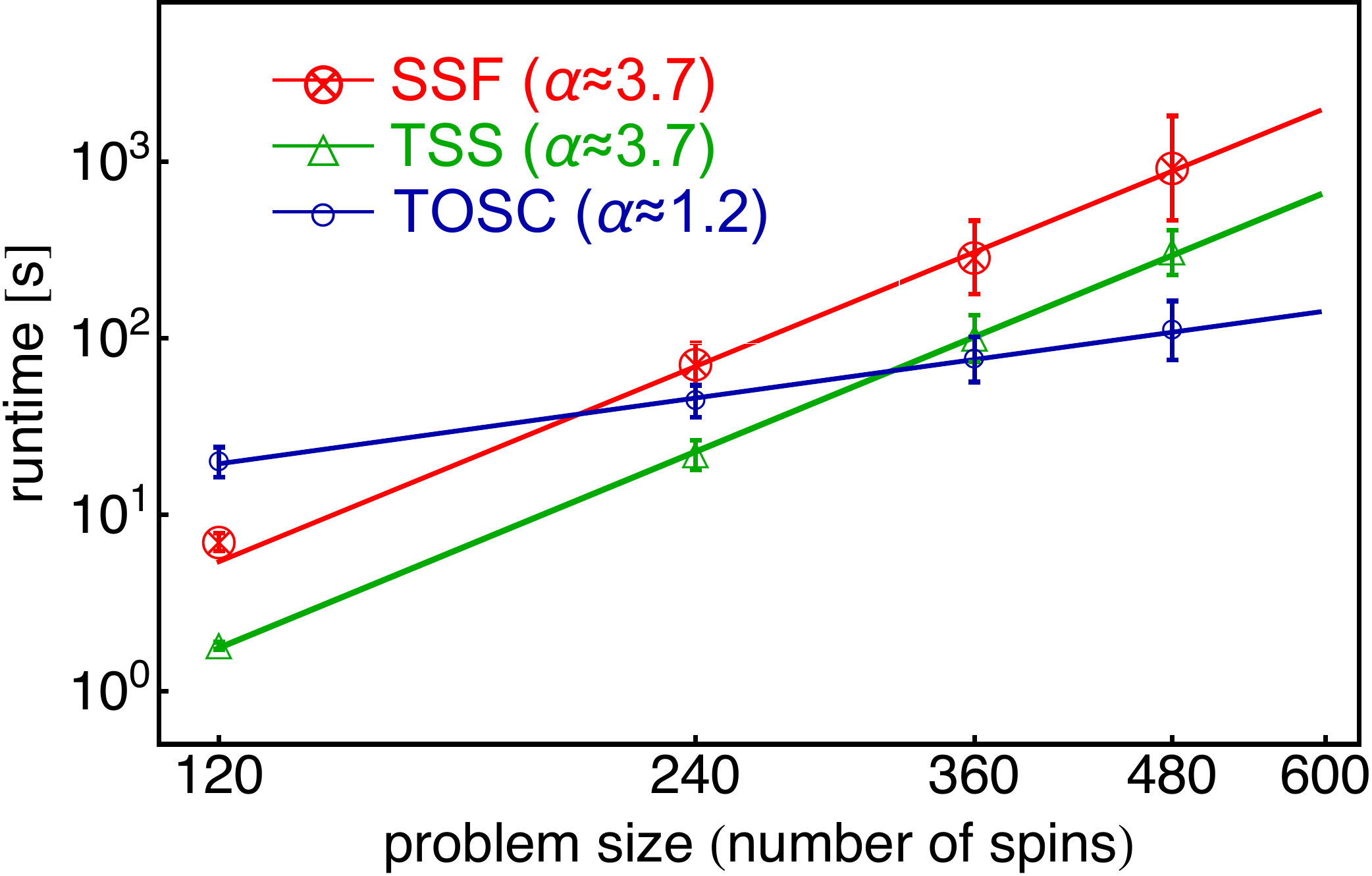}}
   \caption{{\bf Typical runtimes to thermalization of single spin-flip (SSF) parallel tempering, random trees of single spins (TSS) and optimized random trees of spin clusters (TOSC) on random instances of trees of fixed-size clusters.} Runtimes are shown on a log-log scale as a function of problem size. From left to right, the cluster sizes considered are $C=1, 2,$ and $4$. While all three algorithms exhibit power-law scaling, TOSC scales favorably with size as compared to SSF and TSS with a scaling that approaches linear as cluster size grows.}
   \label{fig:resRandomTreesTherm}
\end{figure*}

In Fig.~\ref{fig:resRandomTreesTherm} we plot as a function of problem size the typical `time to thermalization,' which we take here to be the time it takes all PT replicas to traverse the temperature grid from the lowest temperature to the highest and back four times.
Similar to the optimization results of the previous section, we find that typical thermalization runtimes scale polynomially with problem size for all three algorithms. However, while for SSF and TSS the power (the slope) seems to grow with increasing cluster size $C$, the TOSC runtime scales much more mildly, indicating a strong scaling advantage (the errors in the reported powers are of the order of the second decimal place). 

\subsection{Spin glasses on Chimera graphs}

\begin{figure}[ht] 
\includegraphics[width=0.7\columnwidth]{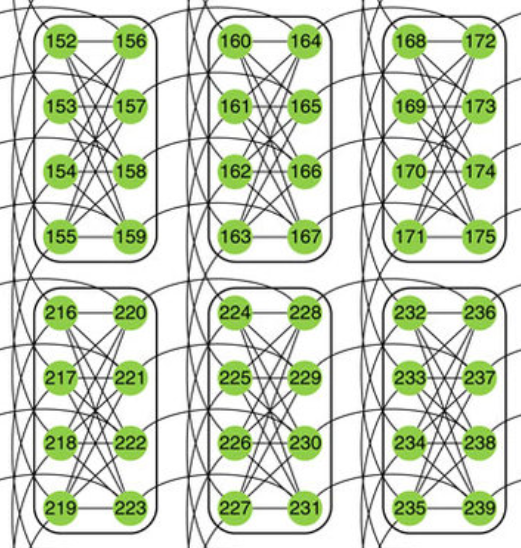}
   \caption{{\bf A 2-cell by 3-cell section of a Chimera graph.} Each cell is a bipartite $K_{4,4}$ graph. Spins are connected either horizontally or vertically to similarly positioned spins in adjacent cells.}
   \label{fig:chimera}
\end{figure}

The problem classes we consider next are random spin glasses whose underlying connectivity graphs are of the Chimera type: two-dimensional arrays of unit cells of eight spins with a $K_{4,4}$ bipartite connectivity~\cite{Choi1,Choi2}. A 2-cell by 3-cell subgraph of a Chimera lattice is shown in Fig.~\ref{fig:chimera}. While the Chimera graph is two-dimensional in nature, it is also non-planar and as such gives rise to difficult spin-glass problems~\cite{barahona:82}. Chimera graphs have become the focus of much attention in recent years in the context of quantum annealing-based optimization due to the commercial availability of prototypical quantum annealing optimizers of spin glasses whose quantum bits are coupled with a Chimera connectivity~\cite{johnson:11,berkley:13,Bunyk:2014hb}. 

\begin{figure*}[ht] 
\subfigure[]{\includegraphics[width=0.67\columnwidth]{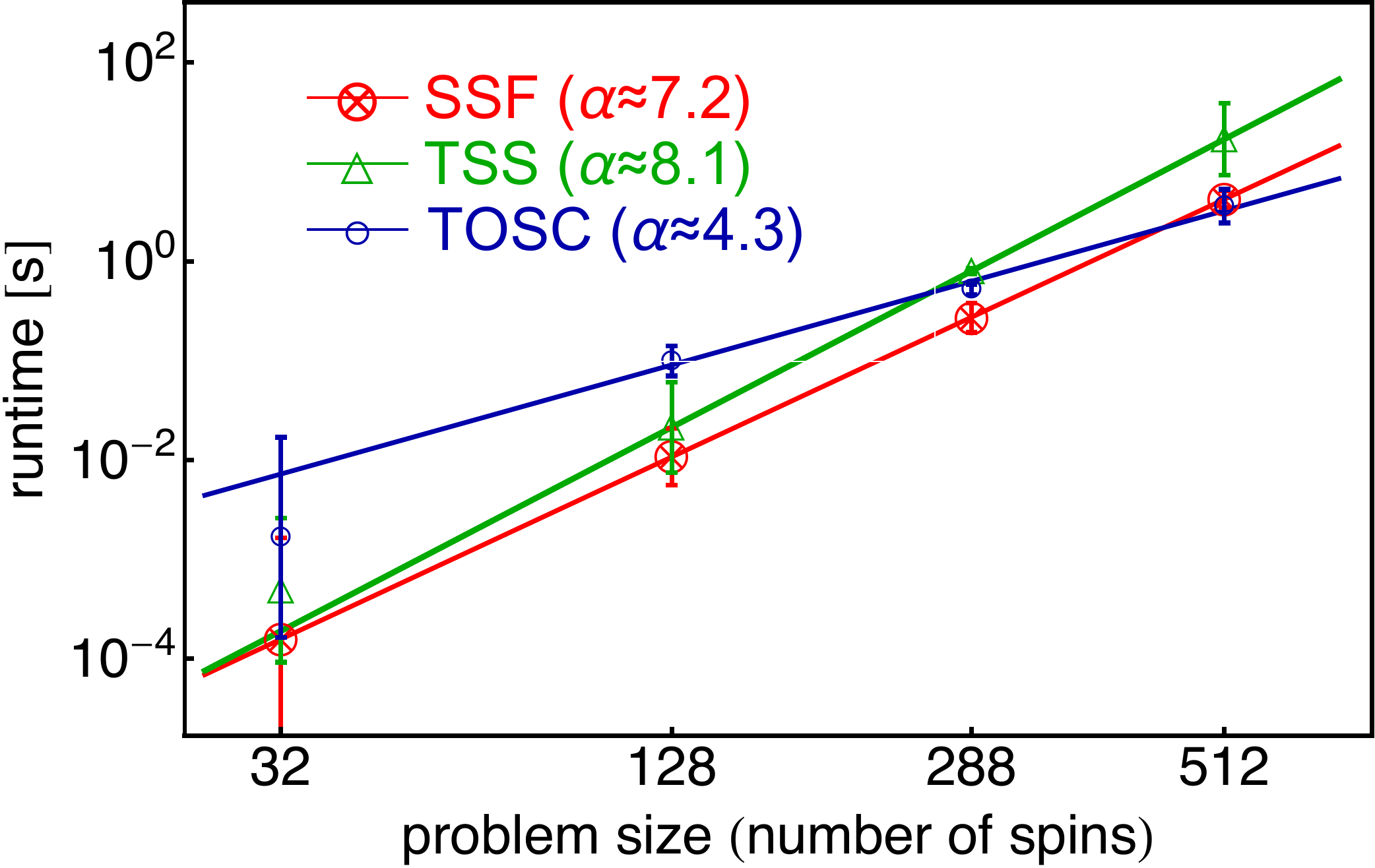}}
\subfigure[]{\includegraphics[width=0.67\columnwidth]{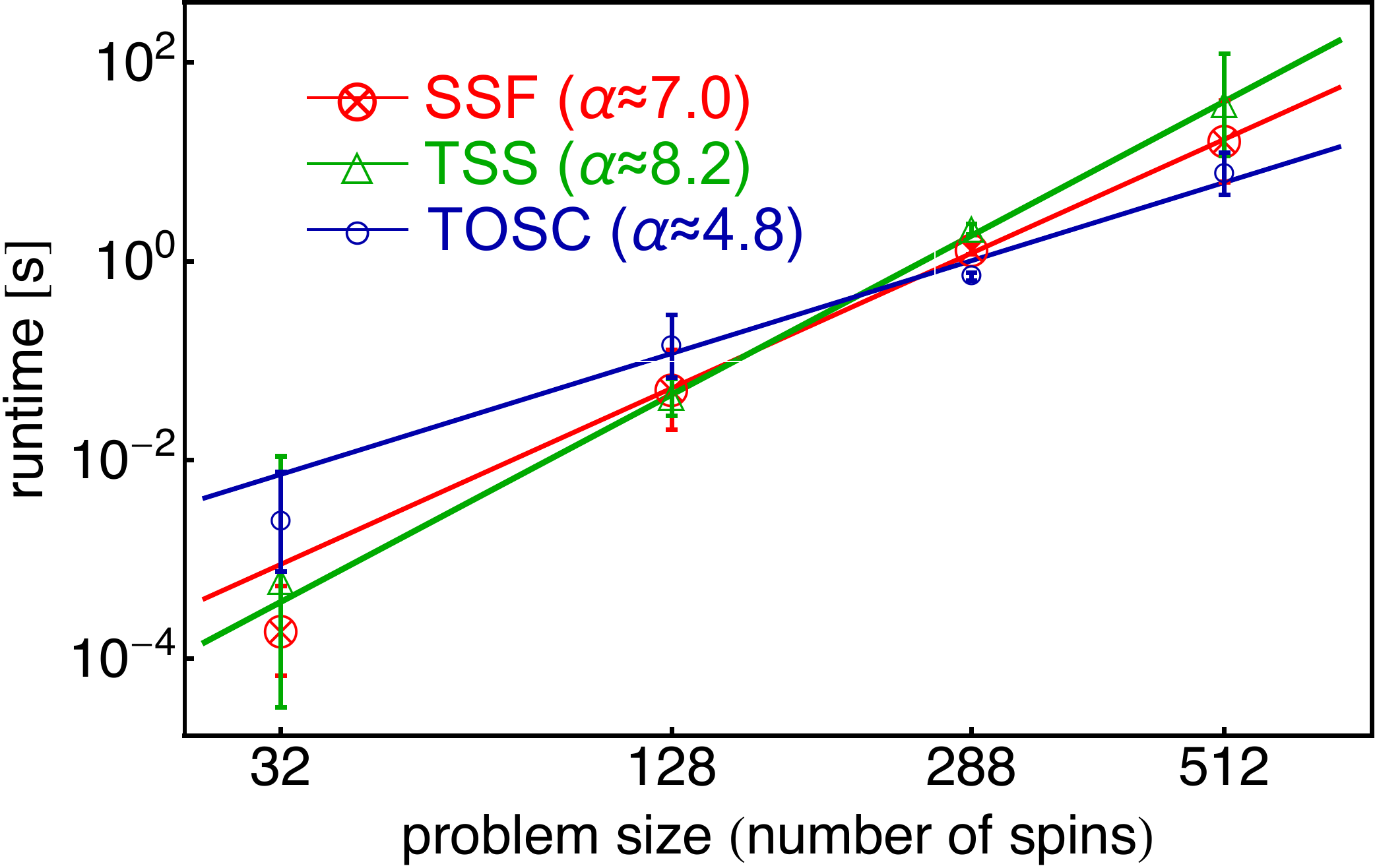}}
\subfigure[]{\includegraphics[width=0.71\columnwidth]{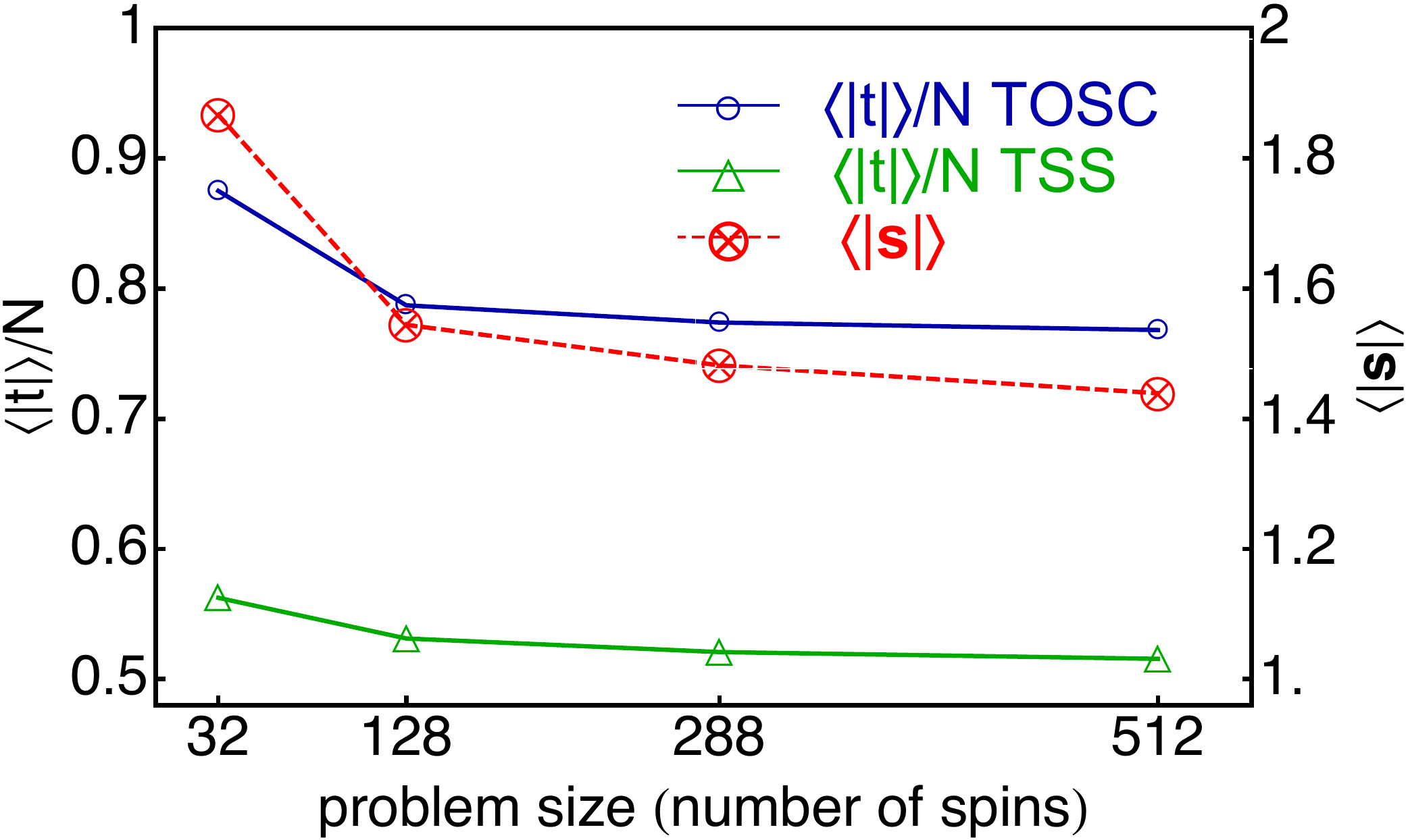}}
   \caption{{\bf Performance of TOSC against SSF and TSS on random Range 1 and Range 3 Chimera instances.} (a) Runtime scaling for Range 1 problems. (b) Runtime scaling for Range 3 problems. (c) Average tree size $\langle |\mathbf{t}|\rangle$ to problem size $N$ as a function of problem size for both TSS and TOSC. Also shown is the average TOSC cluster size $\langle |\mathbf{s}|\rangle$ (dashed line). }
   \label{fig:chimeraResults}
\end{figure*}

Here, we measure the typical runtimes to reach a ground state configuration as recorded by the SSF, TSS and TOSC algorithms on spin glass instances with Chimera connectivities where the couplings are chosen randomly from either $J_{ij} \in \{\pm 1\}$ (`Range $1$' instances) or $J_{ij} \in \{\pm 1, \pm2,\pm3\}$ (`Range $3$'). The results are summarized in Fig.~\ref{fig:chimeraResults}. The first and second panels show the speed-up, exhibited by the milder slope, gained by TOSC as compared to SSF and TSS for both Range $1$ (a) and Range $3$ (b) classes (the errors in the reported slopes are of the order of the second decimal place).  Figure~\ref{fig:chimeraResults}(c) depicts the average coverage of of the trees of TSS (about $50\%$) and TOSC (approximately $75\%$). Also shown is the TOSC average cluster size $\langle |\mathbf{s}|\rangle$.

\subsection{Algorithmic scaling with problem hardness}

We next study the manner in which the performance of the TOSC algorithm correlates with
`instance hardness.' 
Since instances with
large $\tau$ are harder to equilibrate, we follow Refs.~\cite{fernandez:09b,janus:10,fernandez:13,scirep15:Martin-Mayor_Hen,marshall:16}, and use as a measure for the hardness of a random spin glass instance, the mixing time $\tau$ of a single spin-flip PT simulation.  

\begin{figure}[ht] 
\includegraphics[width=0.8\columnwidth]{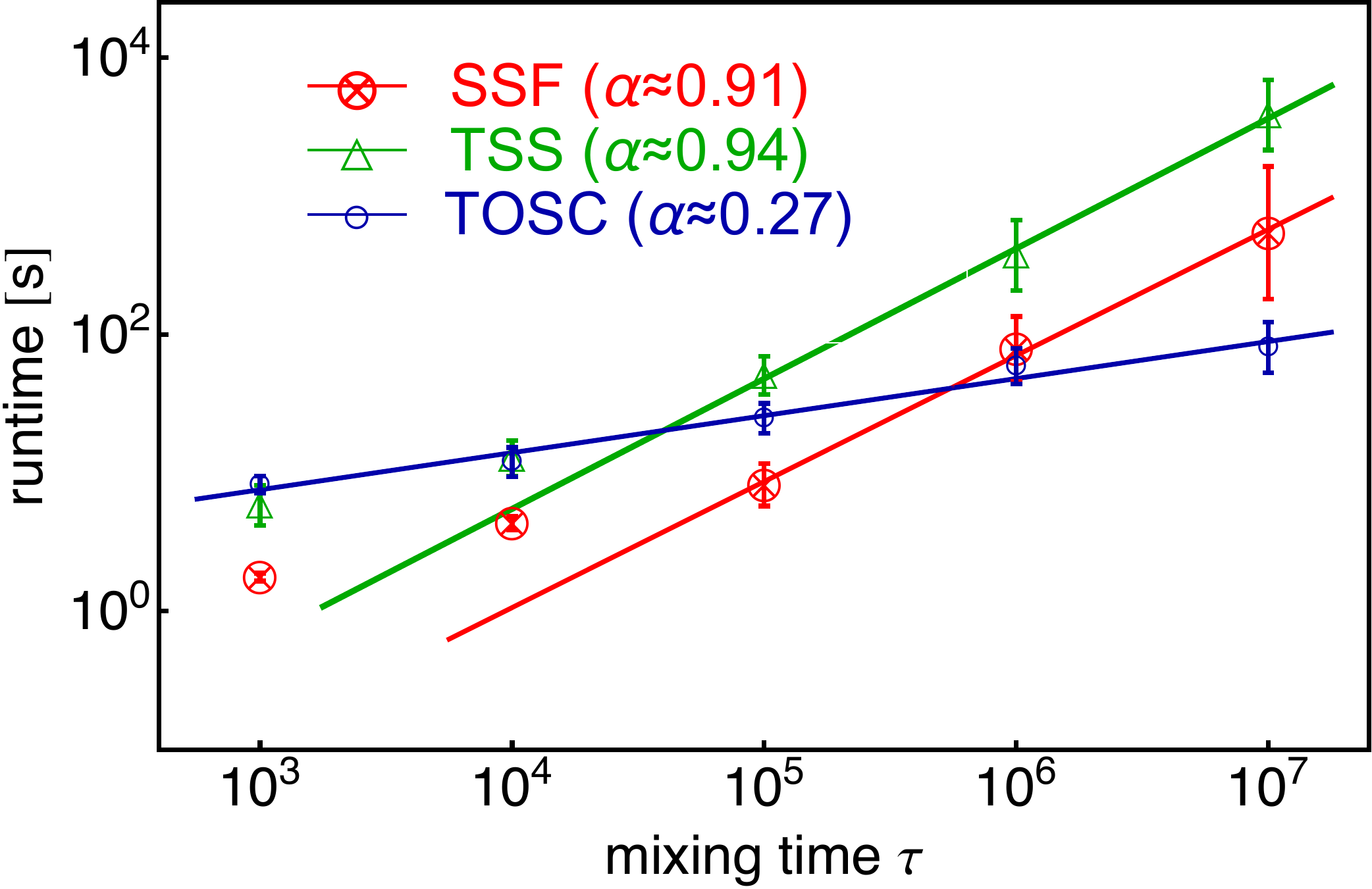}
   \caption{{\bf Algorithmic scaling of TOSC, TSS and SSF on random Chimera instances of size $\mathbf{N=512}$.} While for both  SSF and TSS the runtime scaling with mixing time is approximately linear (with a slope slightly less $1$), the TOSC  scaling is about three times milder, exhibiting a superior robustness against thermal hardness.}
   \label{fig:algorithmic}
\end{figure}

To benchmark the TOSC algorithm, we generate about $10^6$ random instances on an $N=512$-spin Chimera graph and measure the mixing time of each instance~\cite{scirep15:Martin-Mayor_Hen,marshall:16}. As a next step, we group together instances with similar
classical hardness, i.e., similar mixing times, $10^k \leq \tau \leq 3
\cdot 10^k$ for $k=3,4,5,6$ and $7$. For each such `generation' of
$\tau$, we randomly pick $100$ representative instances for the benchmarking of the algorithm
(only 14 instances with $k=7$ were found).  The results are presented in Fig.~\ref{fig:algorithmic} which shows on a log-log scale the typical runtimes of the SSF, TSS and TOSC algorithms as a function of mixing time. 
As is evident, the susceptibility of both SSF and TSS to classical hardness scales approximately linearly with mixing time with a slope slightly below $1$. On the other hand, the TOSC algorithm is found to be much less sensitive to instance hardness, with a slope that is about three times smaller, indicating a clear qualitative advantage over the other algorithms. 

\section{Summary and conclusions\label{sec:discussion}}
We developed a novel algorithm for the optimization and sampling from the Boltzmann distribution of Ising spin glasses of arbitrary sizes and connectivities. The algorithm is designed to take advantage of the connectivity graph of the spin glass by thermalizing randomly generated subgraphs of the input problem---specifically, trees of spin clusters (TOSCs) constructed to optimally balance between the size of the thermalized subgraph and the complexity of doing so. 

Benchmarking the TOSC algorithm against single spin-flip PT and single spin random tree PT, we showed that the TOSC algorithm provides qualitative scaling advantages on all problem classes that have been tested, both in terms of scaling with problem size as well as scaling with instance hardness.
In light of the scaling advantages of the algorithm discussed here and the generality of its scope, we believe that the algorithm will prove to be a useful optimization tool for potentially many classes of problems of practical relevance (once these are cast as spin glasses).

It would be of interest to explore additional and possibly more efficient algorithms for generating random optimal subgraphs based on the figure of merit introduced above, or alternate figures of merit, with which optimal subgraphs are induced and then thermalized on the input problem graph. It is also worth noting that for problem classes that share a single, specific underlying connectivity graph, the generation of globally optimal trees of clustered spins may also be useful, as these may be generated in advance and used in a manner similar to the random structures suggested here.
Furthermore, the TOSC algorithm may also be combined with other efficient approaches in lieu of PT such as the Houdayer algorithm~\cite{Houdayer}.

\begin{acknowledgments}
We thank Victor Martin-Mayor for many insightful discussions on the topics addressed in this paper, and Tameem Albash for useful comments and suggestions.
\end{acknowledgments}

\bibliography{refs}
\end{document}